\documentclass{ieeeaccess}
\usepackage{soul}
\usepackage{cite}
\usepackage{amsmath,amssymb,amsfonts}
\usepackage{algorithmic}
\usepackage{graphicx}
\usepackage{textcomp}
\def\BibTeX{{\rm B\kern-.05em{\sc i\kern-.025em b}\kern-.08em
    T\kern-.1667em\lower.7ex\hbox{E}\kern-.125emX}}
\usepackage{mathtools}
\DeclarePairedDelimiter{\ceil}{\lceil}{\rceil}
\usepackage{textcomp}
\usepackage{caption}
\usepackage{url}
\usepackage{hyperref}
\usepackage{booktabs, tabularx}
\usepackage[referable]{threeparttablex}
\usepackage{siunitx}
\usepackage{stfloats}
\usepackage{subcaption}
\usepackage{qcircuit}
\usepackage{makecell}
\usepackage{float}
\usepackage{braket}
\usepackage{comment}
\newcolumntype{?}{!{\vrule width 2pt}}
\usepackage{multirow}
\usepackage{lipsum}
\usepackage{color,soul}
\usepackage{makecell}
\usepackage[normalem]{ulem}

\usepackage{cleveref}

\newlength\myindent
\usepackage{etoolbox}

\usepackage{pgfplots}
\usepackage{pgfplotstable}
\pgfplotsset{compat=1.7}
\usepackage{tikz}

\NewSpotColorSpace{PANTONE}
\AddSpotColor{PANTONE} {PANTONE3015C} {PANTONE\SpotSpace 3015\SpotSpace C} {1 0.3 0 0.2}
\SetPageColorSpace{PANTONE}

\begin{document}
\history{Date of publication xxxx 00, 0000, date of current version xxxx 00, 0000.}
\doi{10.1109/ACCESS.2024.0429000}

\title{AEQUAM: Accelerating Quantum Algorithm Validation through FPGA-Based Emulation}
\author{Lorenzo Lagostina\authorrefmark{1}, Deborah Volpe\authorrefmark{1},~\IEEEmembership{Graduate Student Member, IEEE}, Maurizio Zamboni\authorrefmark{1}, and Giovanna Turvani\authorrefmark{1}
\address[1]{Department of Electronics and Telecommunications of Politecnico di Torino, Torino, 10129, Italy (e-mail: \href{mailto:lorenzo.lagostina@polito.it}{lorenzo.lagostina@polito.it};
\href{mailto:deborah.volpe@polito.it}{deborah.volpe@polito.it}; 
\href{mailto:maurizio.zamboni@polito.it}{maurizio.zamboni@polito.it};
\href{mailto:giovanna.turvani@polito.it}{giovanna.turvani@polito.it}.)}
\thanks{The authors are with the Department of Electronics and Telecommunications of Politecnico di Torino, Torino, 10129, Italy (e-mail: \href{mailto:lorenzo.lagostina@polito.it}{lorenzo.lagostina@polito.it};
\href{mailto:deborah.volpe@polito.it}{deborah.volpe@polito.it}; 
\href{mailto:maurizio.zamboni@polito.it}{maurizio.zamboni@polito.it};
\href{mailto:giovanna.turvani@polito.it}{giovanna.turvani@polito.it}.)}}

\markboth
{L. Lagostina \headeretal: AEQUAM: Accelerating Quantum Algorithm Validation through FPGA-Based Emulation}
{L. Lagostina \headeretal: AEQUAM: Accelerating Quantum Algorithm Validation through FPGA-Based Emulation}

\corresp{Corresponding author: Giovanna Turvani (email: giovanna.turvani@polito.it).}

\begin{abstract}
This work presents AEQUAM (Area Efficient QUAntum eMulation), a toolchain that enables faster and more accessible quantum circuit verification. It consists of a compiler that translates OpenQASM 2.0 into RISC-like instructions, Cython software models for selecting number representations and simulating circuits, and a VHDL generator that produces RTL descriptions for FPGA-based hardware emulators. The architecture leverages a SIMD approach to parallelize computation and reduces complexity by exploiting the sparsity of quantum gate matrices. The VHDL generator allows customization of the number of emulated qubits and parallelization levels to meet user requirements. Synthesized on an Altera Cyclone 10LP FPGA with a 20-bit fixed-point representation and nearest-type approximation, the architecture demonstrates better scalability than other state-of-the-art emulators. Specifically, the emulator has been validated by exploiting the well consolidated benchmark of mqt bench framework.
\end{abstract}


\titlepgskip=-21pt

\maketitle

\begin{figure*}[t]
	\centering
\includegraphics[width=1\textwidth]{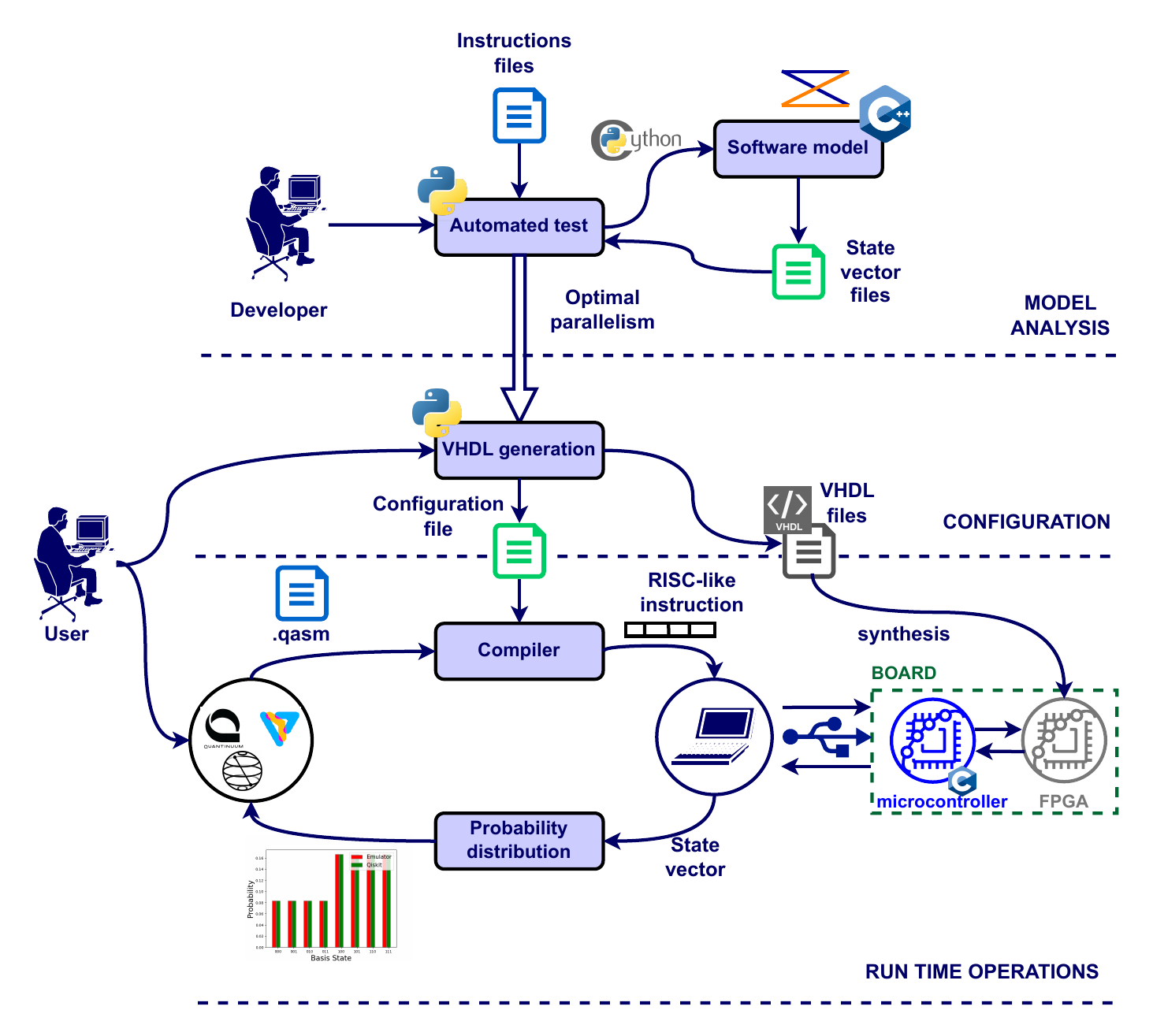}
\caption{\textbf{AEQUAM} (\textbf{Area Efficient QUAntum eMulation}) toolchain for quantum algorithm validation.}
\label{fig:AEQUAM_emulation_toolchain}
\end{figure*}
\section{Introduction}
The industries and researchers' attraction to \textbf{quantum-computing-based solutions} has grown overwhelmingly, pushed by the captivating promise of overcoming the limitations of current classical computers. In particular, the interest is mainly focused on applications like optimization \cite{comboptsurvey,comboptimizationquantumvariational,GroverPatternPerformance,gaspapervlsi}, machine learning \cite{biamonte2017quantum,qmlsurvey}, and chemical simulations \cite{quantumchemistry}, where it is proven that a \textbf{quantum advantage} could be achieved \cite{pirnay2022super,riste2017demonstration,yuan2020quantum}. \\
However, \textbf{validating a new quantum computing algorithm} is challenging, as access to quantum hardware is limited due to high costs, noise, and qubit connectivity constraints. Due to the limited availability and high noise of current quantum hardware, simulation is the most effective way for developing and validating quantum algorithms. Emulators such as AEQUAM allow controlled testing environments and reproducible benchmarks, enabling researchers to verify algorithmic correctness and performance under ideal models.\\
\textbf{Classical simulation} of the \textbf{quantum state evolution} is the most practical method for evaluating the potential of a novel quantum approach since it allows the prediction of the expected ideal behavior with accessible resources. In addition, it also permits access to information that reveals insights into the quantum state, for example, the phase details, which are unavailable with real devices and can be extremely useful for algorithm debugging. Computing the quantum state evolution in software is the most common choice. However, \textbf{software emulation} requires a \textbf{significant amount of time} and \textbf{has substantial memory requirements} \cite{surveyEmulation, memBottleneck}, which limits the dimension of the emulable circuits in terms of qubits and gates. Hardware accelerators, particularly FPGA-based emulators, present a promising alternative by exploiting parallel execution and customized numerical precision to enhance execution speed and efficiency.\\
This article introduces the \textbf{AEQUAM} (\textbf{Area Efficient QUAntum eMulation}) \textbf{toolchain}, an FPGA-based toolchain that accelerates quantum circuit simulation while maintaining a configurable balance between precision and resource usage. As shown in Figure\ref{fig:AEQUAM_emulation_toolchain}, AEQUAM consists of:
\begin{itemize}
    \item Software models that analyze different number representation methods (floating point, fixed-point, and approximations).
    \item A hardware generator that produces customizable FPGA architectures, optimizing qubit count, parallelization, and bit-width representation.
    \item A compiler that translates \textbf{OpenQASM 2.0}  \cite{openqasm2} into \textbf{RISC-like} instructions for hardware execution.
\end{itemize}
The architecture exploits a butterfly selection mechanism, which significantly reduces computational complexity by avoiding redundant matrix operations. The architecture is implemented on an Intel Cyclone 10LP FPGA, demonstrating competitive results in scalability compared to existing solutions.
The key contributions of this work include:
\begin{enumerate}
    \item A novel hardware-oriented quantum simulation approach that balances efficiency and scalability.
    \item A fully automated toolchain for compiling, optimizing, and executing the quantum circuits on FPGA.
    \item A comparative evaluation of AEQUAM against both software-based and hardware-based state-of-the-art emulation methods.
\end{enumerate}
The article is organized as follows. Section \ref{sec:Background} reports the theoretical foundations of the work. In particular, it 
presents the computation required for classical-quantum computer simulation and the related work in the literature. In Section \ref{sec:AEQUAM}, the AEQUAM toolchain is introduced and detailed. Section \ref{sec:architecture} illustrates the emulator architecture.  Section \ref{sec:results} shows the results and explains the validation methodology. Then, Section \ref{sec:discussion} reports the article take home messages. Finally, in Section \ref{sec:conclusions}, conclusions are drawn, and future perspectives are illustrated. 

\begin{figure*}[t]
	\begin{subfigure}[t]{\columnwidth}
	    \centering
\begin{tikzpicture}[line cap=round, line join=round, >=latex]
  \clip(-2.19,-2.49) rectangle (2.66,2.58);
  \draw [shift={(0,0)}, lightgray, fill, fill opacity=0.1] (0,0) -- (56.7:0.4) arc (56.7:90.:0.4) -- cycle;
  \draw [shift={(0,0)}, lightgray, fill, fill opacity=0.1] (0,0) -- (-135.7:0.4) arc (-135.7:-33.2:0.4) -- cycle;
  \draw(0,0) circle (2cm);
  \draw [rotate around={0.:(0.,0.)},dash pattern=on 3pt off 3pt] (0,0) ellipse (2cm and 0.9cm);
  \draw (0,0)-- (0.70,1.07);
  \draw [->] (0,0) -- (0,2);
  \draw [->] (0,0) -- (-0.81,-0.79);
  \draw [->] (0,0) -- (2,0);
  \draw [dotted] (0.7,1)-- (0.7,-0.46);
  \draw [dotted] (0,0)-- (0.7,-0.46);
  \draw (-0.08,-0.3) node[anchor=north west] {$\varphi$};
  \draw (0.01,0.9) node[anchor=north west] {$\theta$};
  \draw (-1.01,-0.72) node[anchor=north west] {$\mathbf {\hat{x}}$};
  \draw (2.07,0.3) node[anchor=north west] {$\mathbf {\hat{y}}$};
  \draw (-0.5,2.6) node[anchor=north west] {$\mathbf {\hat{z}=|0\rangle}$};
  \draw (-0.4,-2) node[anchor=north west] {$-\mathbf {\hat{z}=|1\rangle}$};
  \draw (0.4,1.65) node[anchor=north west] {$|\psi\rangle$};
  \scriptsize
  \draw [fill] (0,0) circle (1.5pt);
  \draw [fill] (0.7,1.1) circle (0.5pt);
\end{tikzpicture}
    \caption{Bloch sphere representation of a qubit.}
    \label{fig:bloch_sphere}
	\end{subfigure}
        \quad
 \begin{subfigure}[t]{\columnwidth}
    \centering
\includegraphics[width=\columnwidth]{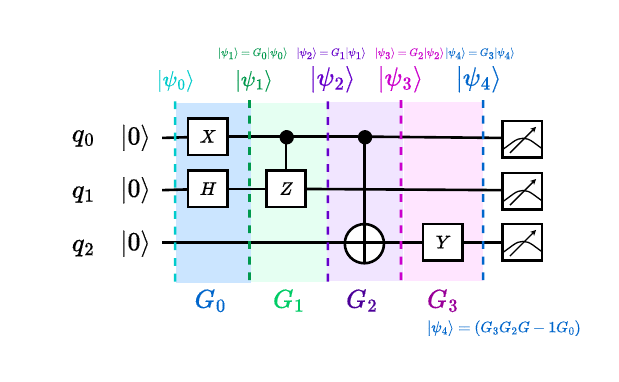}
    \caption{Three-qubit quantum circuit. Vertical dashed lines depict individual layers, described with the matrices $G_x$, each formed as the tensor product of that layer’s gates.}
    \label{fig:QuantumCircuit}
\end{subfigure}	
	\caption{Quantum computing fundamentals.}
	\label{fig:QuantumComputingFundamentals}
\end{figure*}
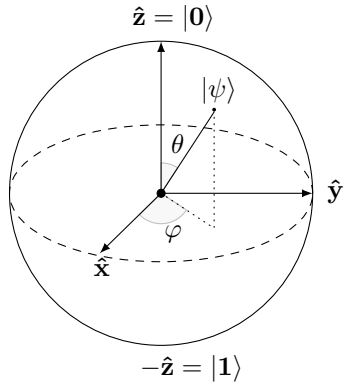
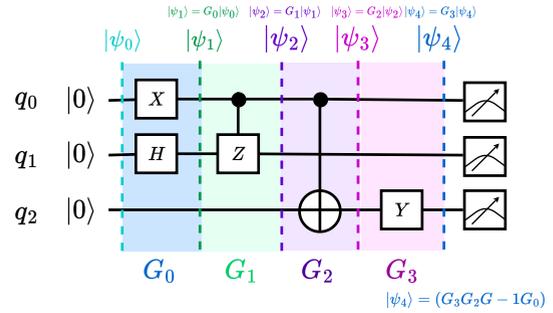

\section{Theoretical foundations}\label{sec:Background}
This section presents the basis of quantum computation,  the essential operations and addresses the challenges in emulating classically quantum state evolution and provides an overview of the related inherent work in the literature,  highlighting the gaps and unmet requirements within this context. A comprehensive overview of quantum computing theory can be found in \cite{nielsen_quantum_2010}.

\subsection{Quantum computing theory}
Quantum computing is a novel computational paradigm that exploits \textbf{quantum mechanics principles}, such as \textbf{superposition} and \textbf{entanglement}, to accelerate data-intensive tasks. To understand the potential of quantum computers over classical computing, it's essential to discuss the key distinctions between the two:
\begin{itemize}
    \item Quantum computing operates within a \textbf{probabilistic paradigm}, where the repetition of the same operations may provide different results, in contrast to the deterministic nature of classical computing.
    \item According to the \textbf{non-cloning theorem}, quantum information cannot be copied, such as the classical one.
    \item While the state of the classical unit of information (bit) can only deterministically assume one of its measurable states (0 or 1), the fundamental unit of quantum information, the \textbf{qubit}, can assume infinite possible
states are given by the linear combination of its basis states, according to the superposition principle.
    \item While classical circuits are physically constructed in a spatial layout, a \textbf{quantum circuit} exists as a time series of transformations applied to the quantum system.
    \item All quantum gates are inherently \textbf{reversible}.
\end{itemize}
In the following, qubit and quantum gate concepts are introduced. 

\subsubsection{Qubit}
The qubit is the fundamental unit of information in quantum computing. Represented using \textbf{Dirac notation}, its state, denoted as $\ket{\psi}$, can be expressed by the \textbf{state vector} \cite{nielsen_quantum_2010}:
\begin{equation}
    \ket{\psi} = a \ket{0} + b \ket{1} = a \begin{pmatrix}1\\0\end{pmatrix} + b\begin{pmatrix}0\\1\end{pmatrix} = \begin{pmatrix}a\\b\end{pmatrix} \, ,
    \label{eq:oneQubitState}
\end{equation}
where $\ket{0}$ and $\ket{1}$ are the basis states 0 and 1, respectively, $a$ and $b$ are complex number called probability amplitudes. \\
This numerical representation shows that a qubit can be in any linear combination of its basis states due to the superposition principle, offering infinite possible states. Its state can be graphically represented as a point on the surface of the \textbf{Bloch sphere}, shown in Figure \ref{fig:bloch_sphere}, by writing it in polar coordinates\cite{nielsen_quantum_2010}:
\begin{equation}
    \ket{\psi} = \cos{\biggl( \frac{\theta}{2} \biggr)} \ket{0} + e^{i \phi} \sin{\biggl( \frac{\theta}{2} \biggr)} \ket{1} \, .
     \label{eq:oneQubitStatePolar}
\end{equation}\\
However, when a qubit is observed (\textbf{measured}), it collapses into either of the two computational bases, $\ket{0}$ and $\ket{1}$. The probability of obtaining these are given by $|a|^2$ and $|b|^2$, respectively.  It is important to notice that these probabilities must satisfy the following relation \cite{nielsen_quantum_2010}: 
\begin{equation}
    |a|^2+|b|^2 = 1 \, .
    \label{eq:probabilityAmplitudeRelations}
\end{equation}
In summary, qubits offer a huge and flexible quantum state space (\textbf{Hilbert} space) with complex probability amplitudes, 
guaranteeing a quantum advantage in data-intensive applications based on the superposition principle.
\begin{figure}
	\centering
\includegraphics[width=\columnwidth]{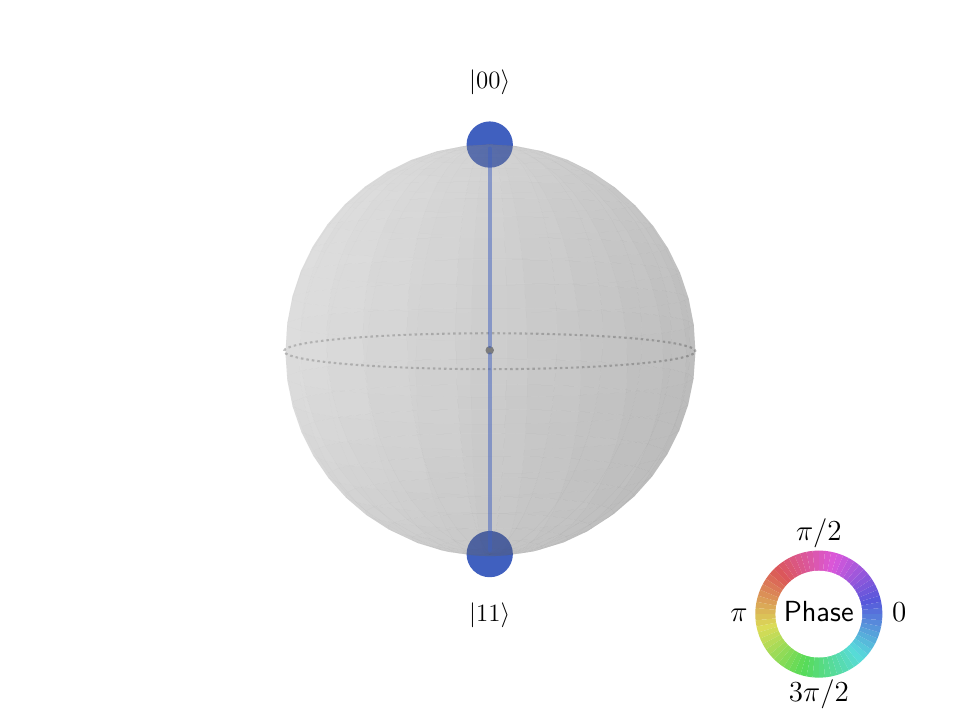}
\caption{Q-sphere representation of a quantum state for a two-qubit quantum circuit. The plot visualizes the state vector on the surface of a sphere, where each point represents a basis state in the quantum superposition. The size of each point indicates the probability amplitude's magnitude, while the color represents the phase. This visualization provides insight into the state’s superposition and phase relationships, demonstrating the entanglement and quantum properties of the circuit.}
\label{fig:QSphere}
\end{figure}
\subsubsection{Quantum circuits}
The numerical representation of a single qubit can be extended  to a $n$-qubit system, defining it as a single state vector $\ket{\psi}$, obtained through the tensor product of the state of the single qubits \cite{nielsen_quantum_2010}:
\begin{align}
\begin{split}
    &\ket{\psi} = \ket{\psi_{n-1}}  \otimes \ket{\psi_{n-2}}  \otimes \cdots \otimes  \ket{\psi_{1}} \otimes \ket{\psi_{0}}  \\
   &= \begin{pmatrix}a_{n-1}\\b_{n-1}\end{pmatrix} \otimes \begin{pmatrix}a_{n-2}\\b_{n-2}\end{pmatrix}  \otimes \cdots \otimes \begin{pmatrix}a_{1}\\b_{1}\end{pmatrix} \otimes \begin{pmatrix}a_{0}\\b_{0}\end{pmatrix} =  \begin{pmatrix}c_{00 \cdots00}\\c_{00 \cdots01}\\ \vdots\\ c_{11 \cdots 10}\\c_{11 \cdots11}\end{pmatrix} \\ &= c_{00 \cdots00} \ket{00 \cdots00} + c_{00 \cdots01} \ket{00 \cdots01} +  \cdots + \\&+  c_{11 \cdots10} \ket{11 \cdots10} + c_{11 \cdots11} \ket{11 \cdots11} \, ,
\end{split}
\label{eq:MultiQubitState}
\end{align}
where the probability amplitude $c_{00 \cdots00}$ is associated with the $ \ket{00 \cdots00}$, $c_{00 \cdots01}$ to $ \ket{00 \cdots01}$ and so forth.\\
For the multiple qubits, the system states can be graphically represented as points on the \textbf{ Q-sphere} surface, as shown in Figure \ref{fig:QSphere}.
\\
The state of a qubit can be modified by applying \textbf{quantum gates}, which can be formally described as unitary $2 \times 2$ matrices. The output quantum state can be mathematically computed as the product between the matrix and the input quantum state \cite{nielsen_quantum_2010}:
\begin{equation}
   \mathcal{U}  \ket{\psi} = \begin{pmatrix} u_{00}  & u_{01} \\ u_{10} & u_{11}\end{pmatrix} \begin{pmatrix} c_0 \\ c_1\end{pmatrix} = \begin{pmatrix} c_0 u_{00} + c_1 u_{01} \\  c_0 u_{10} + c_1 u_{11}\end{pmatrix}  \, ,
\end{equation}
where $\mathcal{U}$ is the unitary matrix of a generic quantum gate. This transformation can be graphically represented as a \textbf{rotation on the Bloch sphere}. \\
Noteworthy single-qubit quantum gates include:
\begin{itemize}
    \item Pauli gates: X, Y, and Z which correspond to the rotation by $\pi$ around $x$, $y$, $z$, respectively.
    \item Hadamard gate H, which creates, starting from the $\ket{0}$ state, the superposition $\ket{+} = \sqrt{\frac{1}{2}} \ket{0} + \sqrt{\frac{1}{2}} \ket{1} $.
    \item Rotational gates $\textrm{R}_x(\theta)$, $\textrm{R}_y(\theta)$, $\textrm{R}_z(\theta)$ and $\textrm{U}_1(\theta)$, which can do rotations by a generic angle $\theta$ on the Bloch sphere.
    \item S, T, $\textrm{S}^{\dagger}$ and $\textrm{T}^{\dagger}$, which performs rotations around the $z$ axis of $\frac{\pi}{2}$, $\frac{\pi}{4}$, $-\frac{\pi}{2}$ and  $-\frac{\pi}{4}$, respectively.
\end{itemize}
To create \textbf{entanglement}, i.e. to establish a strong correlation between qubits where the state of one depends on the state of another, gates involving at least two qubits are required. These gates can be mathematically represented by $2^n \times 2^n$ matrices, where $n$ is the number of involved qubits. In the case of a two-qubit gate \cite{nielsen_quantum_2010}:
\begin{align}
\begin{split}
     \mathcal{U}  \ket{\psi} &= \begin{pmatrix} u_{00}  & u_{01} & u_{02} & u_{03} \\ u_{10}  & u_{11} & u_{12} & u_{13} \\ u_{20}  & u_{21} & u_{22} & u_{23} \\ u_{30}  & u_{31} & u_{32} & u_{33}  \end{pmatrix} \begin{pmatrix} c_0 \\ c_1 \\  c_2 \\ c_3\end{pmatrix} =\\ &= \begin{pmatrix} c_0 u_{00} + c_1 u_{01} + c_2 u_{02} + c_3 u_{03} \\  c_0 u_{10} + c_1 u_{11} + c_2 u_{12} + c_3 u_{13} \\ c_0 u_{20} + c_1 u_{21} + c_2 u_{22} + c_3 u_{23} \\ c_0 u_{30} + c_1 u_{31} + c_2 u_{32} + c_3 u_{33}  \end{pmatrix} \, .
\end{split}
\label{eq:TwoQubitGate}
\end{align}
It is important to emphasize that after the application of a two-qubit gate and the creation of entanglement, the \textbf{qubits cannot be considered as separate entities}, and the state of each one cannot be expressed independently.\\
The most well-known gate in this context is the CNOT gate, a controlled version of the X gate, applied to one qubit (the target) based on the state of another (the control). Its matrix is, if the control qubit is the Least Significant Qubit (LSQ) and the target is the Most Significant Qubit (MSQ), the following \cite{nielsen_quantum_2010}: 
\begin{equation}
    \textrm{CNOT} = \begin{pmatrix} 1  & 0 & 0 & 0 \\ 0  & 0 & 0 & 1 \\ 0  & 0 & 1 & 0 \\ 0  & 1 & 0 & 0  \end{pmatrix} \, .
\end{equation}\\
The term \textbf{quantum circuit} indicates a set of transformations, i.e. quantum gates, applied on a system of qubits. It can be graphically represented as shown in Figure \ref{fig:QuantumCircuit}. 
Computing the state vector evolution of a system of qubits under the application of gates belonging to a quantum circuit is the task of classical simulation.

\subsection{Classical simulation of quantum computers}
Considering that in a quantum circuit, two-qubit gates are commonly applied, the evolution of an $n$-qubit system has to be computed considering the state vector of the overall system, i.e. a vector of dimension $2^n$.  In order to apply a single qubit gate or, more in general, an $m$-qubit gate, where $m < n $, it is necessary to compute the equivalent matrix through the tensor product among gates from MSQ to LSQ of the considered layer, where a non-operation gate, which corresponds to an identity matrix,  is considered where there simply a wire. 
\paragraph*{Example} Considering the circuit in Figure \ref{fig:QuantumCircuit}. The initial state vector is equal to:
\begin{equation*}
    \ket{\psi_0} = \begin{pmatrix}c_{000}\\ c_{001}\\c_{010} \\ c_{011} \\ c_{100}\\ c_{101}\\c_{110} \\ c_{111}  \end{pmatrix} = \begin{pmatrix} 1+0i\\ 0+0i\\0+0i \\ 0+0i \\ 0+0i\\ 0+0i\\0+0i \\ 0+0i  \end{pmatrix} \, .
\end{equation*}
 In the first layer, an X gate is applied to qubit 0 and an H gate is applied to qubit 1. The equivalent matrix for the layer can be computed as follows:
 \begin{equation*}
   \textrm{G}_1 =  \mathbb{I}  \otimes \textrm{H} \otimes  \textrm{X} = \begin{pmatrix}1 & 0 \\ 0 & 1  \end{pmatrix} \otimes \frac{1}{\sqrt{2}} \begin{pmatrix}1 & 1 \\ 1 & -1  \end{pmatrix} \otimes  \begin{pmatrix}0 & 1 \\ 1 & 0  \end{pmatrix}\, .
\end{equation*}
The state vector after the application of the first layer is equal to:
 \begin{equation*}
   \ket{\psi_1} = \textrm{G}_1 \ket{\psi_0}\, .
\end{equation*}
Analogously, for computing the second layer equivalent matrix, it is necessary to perform the following operations: 
\begin{equation*}
   \textrm{G}_2 =  \mathbb{I}  \otimes \textrm{CZ}_{0,1} \, ,
\end{equation*}
and the output state can be computed as:
 \begin{equation*}
   \ket{\psi_2} = \textrm{G}_2 \ket{\psi_1}\, .
\end{equation*}
The presented procedure will be repeated for each layer until the final state is obtained. 

\vspace{0.8cm}
\begin{figure*}[t]
	\centering
	\includegraphics[width=0.9\textwidth]{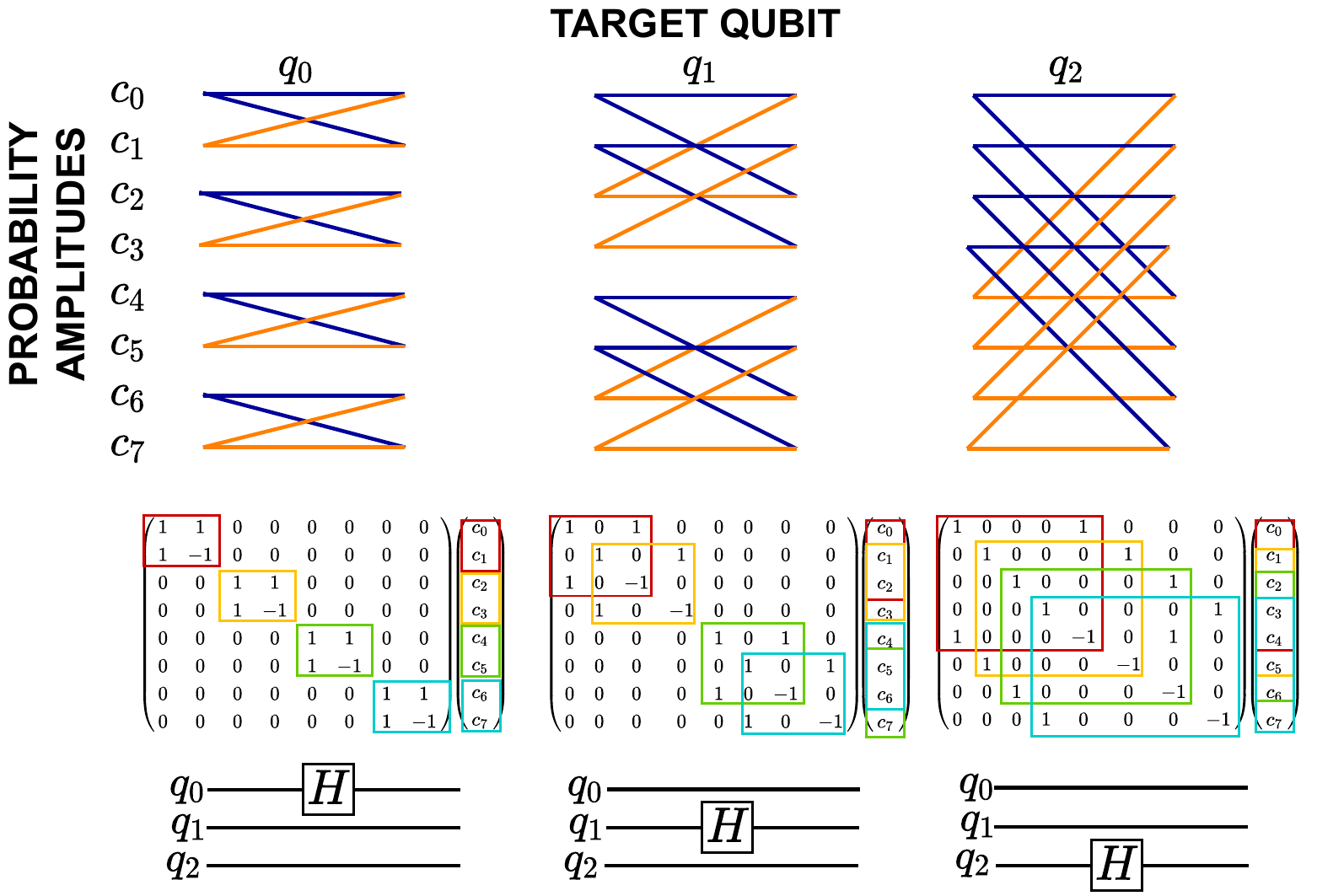}
\caption{The butterfly-like mechanism for selecting interacting couples of probability amplitudes for each potential target qubit in a three-qubit system. As an example, the equivalent matrix of a three-qubit circuit is reported in which the Hadamard gate is applied to each of the qubits. }
\label{fig:butterfly}
\end{figure*}
Alternatively, it is possible to compute the equivalent matrix of the entire circuit and apply it to the initial state to obtain the final state directly. The equivalent matrix of the overall circuit can be obtained by multiplying the equivalent layer matrices from the last to the first, as in the following:
 \begin{equation*}
   \textrm{G} = \textrm{G}_{L} \textrm{G}_{L-1} \cdots  \textrm{G}_{2} \textrm{G}_{1} \, ,
\end{equation*}
where $L$ is the number of layers and $\textrm{G}_{i}$ is the equivalent matrix of the $i^{\textrm{th}}$ layer.\\
Unfortunately, the complexity of operations required for simulating a quantum circuit grows exponentially with the number of qubits involved as the memory requirements, making this task crucial.  In fact, the main challenge in this research field is to limit the exponential increase of resources, both in terms of memory and computation. 

\subsection{Related works}
In recent years, various FPGA architectures have been proposed to overcome the limits of software emulation.\\
Most hardware emulators in the literature are based on parallel computing of the matrix-vector product layer by layer until the final state is computed. For example, \cite{pilch2019fpga} proposed a universal and scalable quantum computer emulator, which loads from a processor the equivalent layer matrix to be applied and computes the new state by parallelizing the product with the state vector, such that each layer can be evaluated in a single clock cycle. The main problem with this approach is scalability since the area occupied by both computation and memory increases vertiginously with the number of qubits. Indeed, considering an Intel Cyclone V (300k logic elements), they implement only two qubits.\\
The parallel matrix-vector product methodology is also employed in reference \cite{mahmud2018scalable}, with a floating-point number representation instead of a fixed-point one. However, this approach further limits the number of emulated qubits per logic element. In fact, considering an Intel Arria 10 FPGA with 10AX115N4F45E3SG (1150k logic elements), the maximum circuit size is restricted to four qubits. 
The scalability of the system was enhanced in \cite{mahmud2020efficient} by relocating the state vector storage from the FPGA to an external memory. However, inserting an off-chip memory inside the computational loop drastically reduces the maximum operating frequency of a system. Additionally, the hardware was optimized for a unique
algorithm, the quantum Fourier transform, resulting in a 16-qubit emulator.\\
To minimize unnecessary operations, particularly zero products due to the sparse nature of gate matrices, and to bypass the computation and storage of layer matrices, a processor-based architecture was proposed in \cite{Reis}. This architecture executes only the essential operations in parallel, employing a butterfly selection mechanism outlined in \cite{negovetic2002evolving}. While this approach is fascinating and shares similarities with AEQUAM hardware, this architecture has limitations. It supports only a limited set of quantum gates—specifically, Pauli X, CNOT, Toffoli, and Hadamard. 

\section{The AEQUAM toolchain} \label{sec:AEQUAM}
\textbf{AEQUAM} (\textbf{Area Efficient QUAntum eMulation}) toolchain is a framework for improving the \textbf{speed} and \textbf{accessibility} of \textbf{quantum algorithm validation}. The main idea behind AEQUAM is to provide a \textbf{user-friendly reconfigurable emulator}, exploitable for both teaching and research purposes, with a \textbf{reduced computational complexity} obtained by exploiting a \textbf{butterfly-like mechanism}.  \\
As shown in Figure \ref{fig:AEQUAM_emulation_toolchain}, it is composed of a \textbf{compiler} --- which allows the translation of openQASM 2.0 in emulator-compatible instructions ---, \textbf{software models} --- exploited for studying the impact of number representation on the final results  ---, and a \textbf{hardware description generator} --- providing the RTL description of the emulator \textbf{architecture} supporting the desired number of qubits for the synthesis on the target FPGA ---, which are described in detail in this section. \\
Compared to traditional software emulation frameworks, AEQUAM achieves a notable reduction in memory usage, thanks to short fixed point number representation, and execution latency for small to medium-sized quantum circuits. Its hardware-aware design allows efficient usage of logic resources and power. However, the trade-off comes in terms of scalability, as the number of concurrently emulated qubits is constrained by available on-chip memory and logic slices. AEQUAM is thus particularly effective in constrained, real-time environments or for developing embedded quantum applications.\\
The toolchain AEQUAM is publicly available on \href{https://drive.google.com/drive/folders/1sbn9oUAO3Rgr_xAzV4Cf5xwtV8WIm61u?usp=sharing}{GitHub}. 

\subsection{Butterfly-like mechanism} \label{sec:butterflyMechanism}
The \textbf{butterfly-like} selection mechanism of \textbf{interacting couples in the state vector}, proposed for the first time in \cite{negovetic2002evolving}, \textbf{reduces the computational complexity} in quantum circuit simulations. 
In particular, it exploits the \textbf{sparse nature} of the equivalent gate matrices, as shown in the example in Figure\ref{fig:butterfly}. In this way, it avoids unnecessary operations and the computation of the equivalent layer gate matrix. The pattern of interacting couples mirrors that of the \textbf{Fast Fourier Transform} (\textbf{FFT}) \cite{nussbaumer1982fast} butterfly scheme, from which it derives its name. This pattern can be identified by noting that the equivalent circuit matrice of a single-qubit gate applied to a target qubit is composed of $2^{N-1}$ kernels, where $N$ is the number of circuit qubits (highlighted with a different color in the examples). Each kernel is equal to the original single-qubit gate matrix, acting only on two probability amplitudes of the state vector (a couple). The positions of these blocks, and consequently, the interacting couples change as the target qubit changes with the butterfly pattern (Figure \ref{fig:butterfly}).

\begin{figure}[t]
	\centering
	\includegraphics[width=\columnwidth]{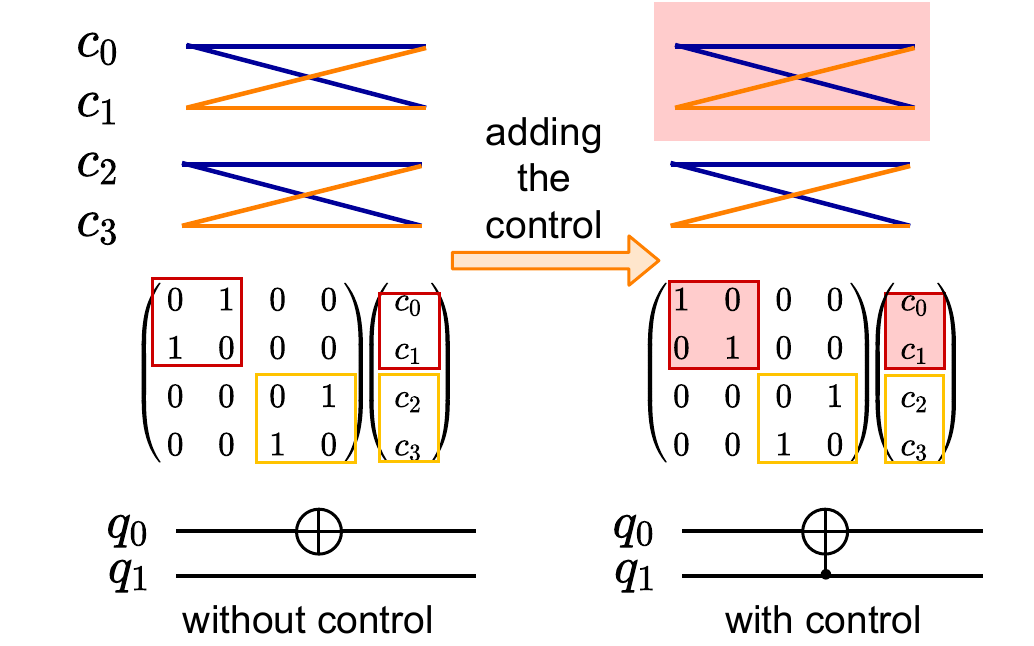}
\caption{The butterfly-like mechanism for selecting interacting couples of probability amplitude for each potential target qubit in a two-qubit system, considering the application of a controlled gate, implying that only a specific subset of couples has to be considered. In particular, in the reported example, the first couple (with a red background) has to be ignored, since the matrix applies an identity gate on it.  }
\label{fig:butterfly_control}
\end{figure}

In the case of a two-qubit control gate, half of the blocks turn into an \textbf{identity matrix}, as shown in Figure 4. The application of identities on probability amplitude couples leaves them unchanged, resulting in redundant operations that can be skipped. These couples can be identified since they involve the basis state with the control qubit equal to 0.  Therefore, it is possible to conclude that two-qubit control gates can be defined as a special case of their single-qubit counterpart, where only couples with the control qubit at 1 are taken into account. For example, in Figure \ref{fig:butterfly_control}, the probability amplitudes $c_{\ket{00}}$ and $c_{\ket{01}}$ are left unchanged since they are associated with $\ket{00}$ and $\ket{01}$ basis states, which present the control qubit (i.e. the MSB in this case) equal to 0.

\subsection{Compiler}\label{sec:compiler}
\begin{figure}[h]
    \centering
    \includegraphics[width=1\columnwidth]{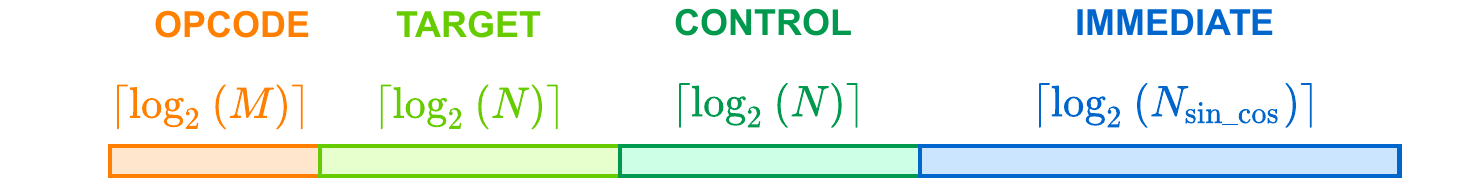}
    \caption{RISC-like instruction format.}
    \label{fig:AEQUAM_operations_format}
\end{figure}

\begin{figure*}[t]
	\centering	\includegraphics[width=1\textwidth]{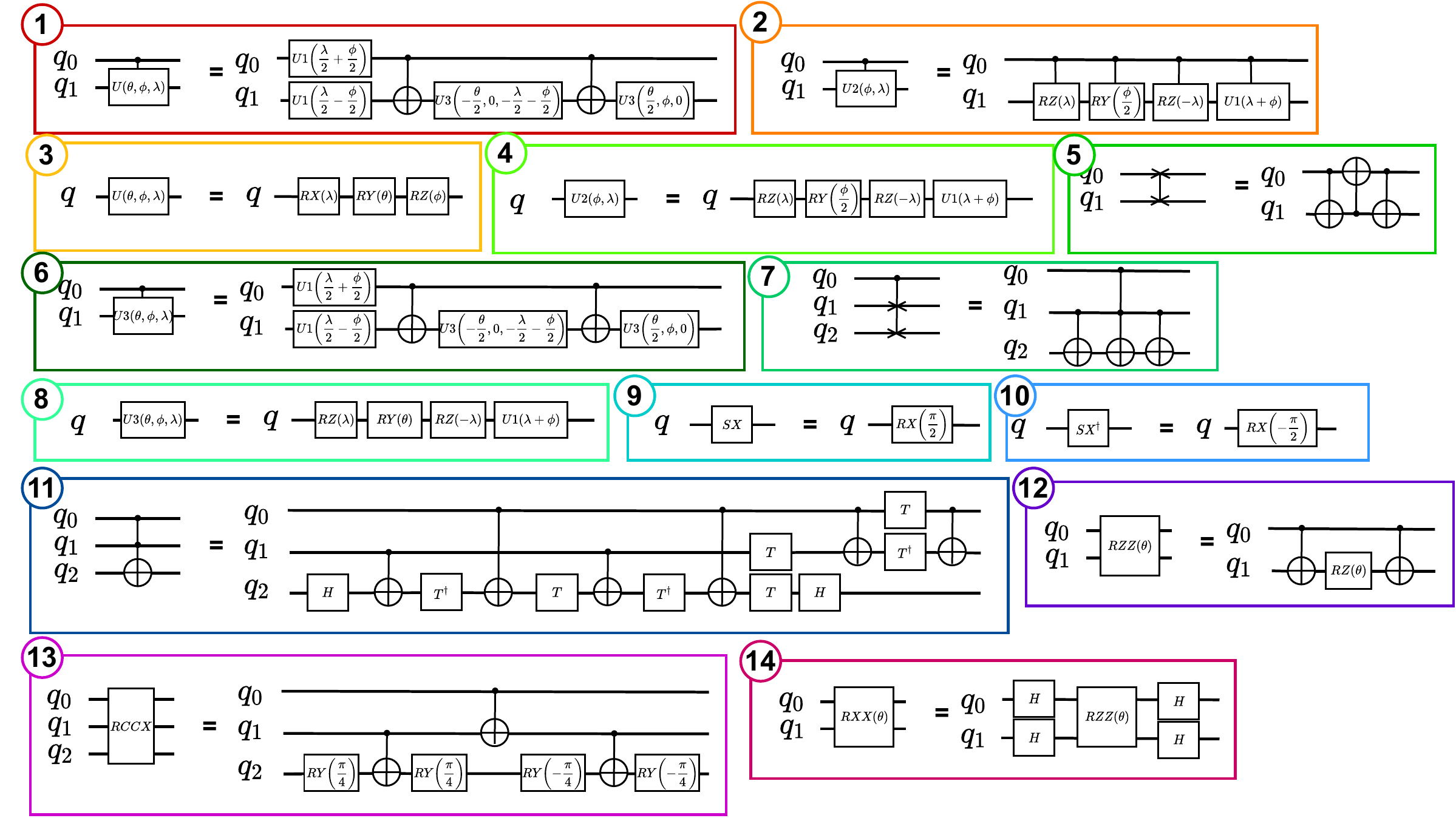}
\caption{Gates' equivalences.}
\label{fig:GateEquivalence}
\end{figure*}

To fully leverage the butterfly-like mechanism and implement it efficiently in hardware, the sequence of gates of a quantum circuit has to be written as a set of \textbf{architecture-specific instructions}. The best instruction format,  considering crucial gate information for execution --- the position of the target, the position of the eventual control, the type of gate, and angle-related data for rotational gates ---is the one shown in Figure \ref{fig:AEQUAM_operations_format}, similar to that of RISC processors. To ensure compatibility with leading quantum frameworks, such as \textbf{Qiskit}, \textbf{PennyLane}, and $\textrm{t} \vert \textrm{ket} \rangle$ and facilitate potential users, the compiler, implemented for obtaining the emulator instructions, requires as input the quantum circuit description in \textbf{OpenQASM 2.0} \cite{OPENQASM}, that can be generated and handled by all main quantum tools.
\\
As it is possible to notice, the RISC-like instruction includes an \textbf{opcode}, identifying the gate to execute, bits defining the \textbf{target} and \textbf{control} qubits (for a single-qubit gate, target, and
control field coincides), and an \textbf{immediate} field containing angle information essential for rotational gate execution. To avoid an excessive area enhancement due to the insertion of logic for trigonometric functions evaluation, in this preliminary version of the toolchain, the sine and cosine of the parametric angle $\theta$, necessary for rotational gates, are pre-computed at the compilation level and stored in a proper array. Therefore, the immediate instruction corresponds to the position of the sine and cosine couple in the array. In order to save space in the sine-cosine array, the compiler can recognize if the angle of a gate was already evaluated for a previous gate of the quantum circuit. In this case, the angle evaluation is avoided, and the immediate field is assigned to its sine and cosine position in the array.\\
The length of each field, excluding the gate opcode, depends on the architecture configuration, which is specified through a proper file to the compiler as input. The target and control qubit fields are $\ceil{\log_2{(N_\textrm{q})}}$ long, where $N_\textrm{q}$ is the maximum number of qubits emulable by the synthesized architecture. The immediate field's length depends on the memory size allocated in the architecture specified in the configuration file. It allows users to tune its value based on the characteristics of the circuits they aim to simulate. For example, for variational circuits, a long array is required since they involves several rotational gates whose angles can differ substantially. In contrast, for Grover Search circuits, a very short array is sufficient, while the angle involved in creating the phase oracles is the same in each rotation, substantially reducing the number of different angles effectively involved in the circuit. This computation approach is more effective in the second case. 
\begin{table}[h]
    \centering
    \caption{Opcode and matrix of the supported gates.}
    \begin{tabular}{?c|c|c?}
    \noalign{\hrule height 1.5pt}
         \textbf{Gate} & \textbf{Opcode} & \textbf{Matrix}\\ \noalign{\hrule height 1.5pt}
         $X$ & 0000 & $\begin{pmatrix} 0  & 1 \\ 1 & 0\end{pmatrix} $ \\  \hline
         $Y$ & 0001 & $\begin{pmatrix} 0  & -i\\ i & 0\end{pmatrix}$ \\ \hline
         $Z$ & 0010 & $\begin{pmatrix} 1  & 0 \\0 & -1\end{pmatrix} $   \\ \hline
        $H$ & 0011 & $\frac{1}{\sqrt{2}}\begin{pmatrix} 1  & 1 \\ 1 & -1\end{pmatrix} $   \\ \hline
        $S$ & 0100 & $\begin{pmatrix} 1 & 0\\ 0 & i\end{pmatrix}$   \\ \hline
        $S^{\dagger}$ & 0101 & $\begin{pmatrix} 1  & 0 \\ 0 & -i\end{pmatrix} $  \\ \hline
                $T$ & 0110 & $\begin{pmatrix} 1  & 0 \\ 0 & e^{i\frac{\pi}{4}}\end{pmatrix}$   \\ \hline
        $T^{\dagger}$ & 0111 & $\begin{pmatrix} 1  & 0 \\ 0 & e^{-i\frac{\pi}{4}}\end{pmatrix}$  \\ \hline
        $RX$ & 1000 & $\begin{pmatrix} \cos{(\frac{\theta}{2})}  & -i\sin{(\frac{\theta}{2})}\\ -i\sin{(\frac{\theta}{2})}   & \cos{(\frac{\theta}{2})}\end{pmatrix}$   \\ \hline
        $RY$ & 1001 & $\begin{pmatrix} \cos{(\frac{\theta}{2})}  & -\sin{(\frac{\theta}{2})}\\ \sin{(\frac{\theta}{2})}   & \cos{(\frac{\theta}{2})}\end{pmatrix}$    \\ \hline
         $RZ$ & 1010 & $\begin{pmatrix} e^{-i\frac{\theta}{2}}  & 0\\ 0   & e^{i\frac{\theta}{2}}\end{pmatrix}$   \\ \hline
          $U1$/$P$ & 1011 & $\begin{pmatrix}1  & 0 \\ 0 & e^{i\theta}\end{pmatrix}$  \\
 \noalign{\hrule height 1.5pt}
    \end{tabular}
    \label{tab:GateOpcode}
\end{table}\\
The gate opcode requires four bits for any configuration since the architecture supports twelve different gates. Table \ref{tab:GateOpcode} shows the gate-opcode association. AEQUAM also supports the other gates supported by OpenQASM 2.0 by exploiting \textbf{gates' equivalences} shown in Figure \ref{fig:GateEquivalence}. 
\\
Another feature supported by the compiler is the management of \textbf{user-defined gates}. \\
In the current version of the toolchain, the measurement operation is not implemented at the hardware level but is eventually obtained in software (as described in Section \ref{sec:hardware}). This choice is related to the principal target of the project, i.e., facilitating functional verification and algorithmic exploration. Therefore, generating the state vector as an output, rather than performing direct measurements, provides more comprehensive insights into the simulated quantum circuit, including phase information otherwise inaccessible with real quantum hardware.  Moreover, emulating the measurement in hardware would be area-expensive since it requires a random number generator \cite{BAKIRI2018135} employed exclusively for this operation, which usually occurs only once at the end of the computation. Furthermore, the time and area overhead required for evaluation are significant, substantially reducing the approach's advantages. This aspect is beyond the scope of this work, as for debugging and validating new quantum algorithms, the state vector --- which contains all the information about the quantum state --- is more useful than the measurement distributions.  
Consequently, the measure command in the OpenQASM 2.0 is ignored at compilation time. \\
As a result, the \texttt{if} command of OpenQASM 2.0, which allows the execution of different gates based on the result of a qubit measurement, is currently not supported. However, future releases will incorporate support for this feature by partitioning the OpenQASM whenever an \texttt{if} command is encountered, enabling software-based measurements and executing the second circuit contingent on the measured outcome.
\begin{figure*}[t]
	\centering	\includegraphics[width=0.8\textwidth]{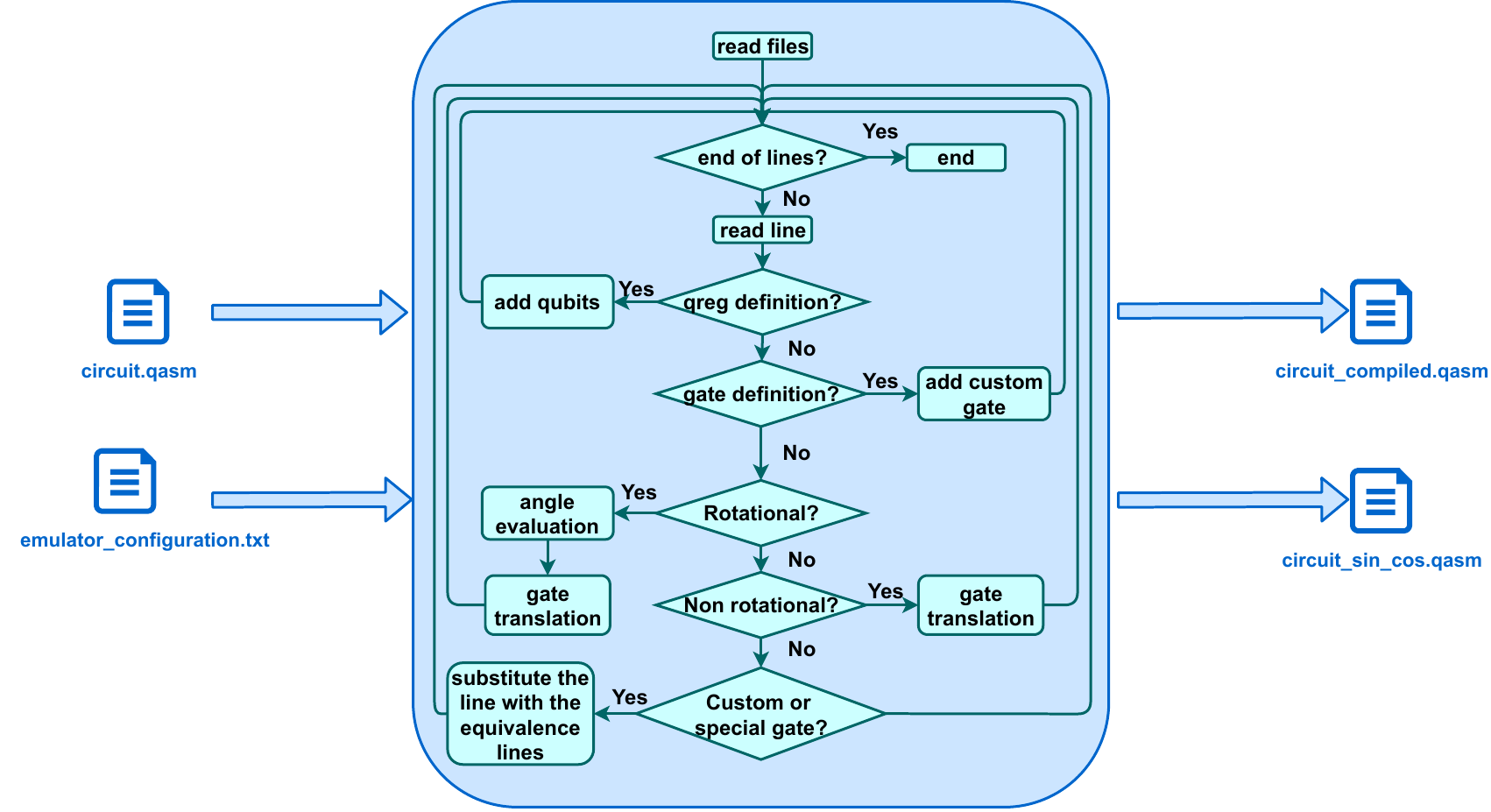}
\caption{Compiler structure}
\label{fig:compiler}
\end{figure*}
The compilation steps required to obtain the architecture-specific instructions from the OpenQASM description are resumed in Figure \ref{fig:compiler}. In particular, the compiler requires as input the \texttt{.qasm} file of the circuit and the architecture configuration file. The latter provides information on the number of supported qubits and the immediate length, while the former is processed line by line. When a new quantum register is defined in a line, the associated qubits are incorporated into the circuit's qubit dictionary.
Similarly, if a custom gate is defined, its translation is memorized in a proper dictionary. Each supported gate is translated by identifying the target and, eventually, the control one. For rotational gates, sine and cosine values are computed unless already present in the sine-cosine list or their position in the array is determined. If a user-defined gate or a special gate, i.e., one of the defined equivalences, is identified, the line is replaced with the gate translation, substituting the parameters.   \\
The compilation outcome is a file containing the compiled instructions, with the first line specifying the number of effective qubits employed and another containing the list of the sine and cosine, with the first line specifying the number of effective sine-cosine couples. The files can be written in binary or employing integer numbers. \\

\subsection{Software models}\label{sec:SoftwareModels}
In order to validate the conceptual foundation of the proposed architecture and estimate the impact of parallelism and result approximation, the toolchain includes software models. These models are implemented in \href{https://cython.org}{Cython} language, i.e., with a \textbf{C++} core, guaranteeing an efficient execution of the most complex computational part, and an external \textbf{Python interface}, which allows compatibility with the compiler and with the main quantum frameworks. They covered \textbf{different number representations and approximation methods} (floating point, fixed-point with truncation, fixed-point with nearest, and fixed-point with nearest even) \cite{koren2018computer}. \\
The \textbf{floating point model} was developed to verify the effectiveness of the butterfly selection mechanism -- without limitations in terms of results accuracy related to fixed-point number representation  ---, but it can also be employed alone as an \textbf{alternative software simulator} to those provided by the main quantum frameworks since it is quite efficient. Indeed, the C++ core guarantees \textbf{higher performance} than a pure Python implementation, and the butterfly mechanism substantially \textbf{reduces computational complexity} by executing only the necessary operation.  
Unfortunately, due to the limited resources of the devices on which they were executed, the software implementation is not parallelized using multi-thread or multiprocessing programming, as required for making the most of the butterfly mechanism. However, the C++ core of the simulator would permit the parallelization with a limited amount of changes in the code, e.g., by adding, for example, the  \texttt{\#pragma omp for} option for the gate execution on the couples.\\
Fixed-point models allow the declaration of the wanted precision for the number representation to determine the best hardware design. Typically, fixed-point representation offers a more cost-effective design in terms of area and latency. In this implementation, numbers are represented as standard integers, with the Least Significant Bit (LSB) having a virtual weight of $2^{-(N_{\textrm{parallelism}}-2)}$, where $N_{\textrm{parallelism}}$ is the total number of bits considered for number representation. In fact, considering that the probability amplitudes of the state vector have to be in the range $[-1, 1]$ to not violate the probabilities rules mentioned in Section \ref{sec:Background}, the number of integer bits in fixed-point representation can be set to two a priori. However, choosing the correct number of decimal bits is crucial for achieving the best possible design, so the model's computational precision can be set to span among the possible values and make a conscious choice. \\
Following each complex numerical operation, such as multiplication, since it provides the results on more bits than input, a bit reduction is necessary to maintain a constant bit representation. This operation, involving a loss of information, is influenced by the chosen rounding strategy. Truncation, a simple mechanism implemented through right-shift operations, is a possible approach. 
Another strategy is the nearest approximation, representing the numerical value with the nearest available value in the representable range.  The third option, nearest even, refines the standard nearest strategy, balancing situations where the value falls precisely halfway between two possible rounded values to minimize approximation errors. \\
As in the proposed architecture, each supported gate is executed in each software model by computing only the required operations. As it is possible to notice, the gates can be virtually subdivided into three categories: 
    
\begin{itemize}
\item Gates that require only sign inversions and position of elements in the register file ($X$, $Y$, $Z$, $S$ and $S^\dagger$).
\item Gates implementable through an addition and a multiplication ($H$, $T$ and $T^\dagger$).
\item Gates executable with addition and two multiplications ($RX$, $RY$, $RZ$ and $U1$) and using sine and cosine of a generic angle (rotational gates).
\end{itemize}

\subsection{Proposed architectures} \label{sec:architecture}
The emulator architecture is inspired by the RISC processor for two main reasons. The RISC, which works on a reduced instruction set, aligns point current quantum computers, which employ only a limited number of gates. Moreover, a possible future perspective of the current project could evolve into a hardware accelerator to integrate it into a real RISC file. \\
Therefore, the architecture includes a \textbf{register file}, storying the real and imaginary parts of the state vector (\textbf{Quantum State Register File}), a decode unit for interpreting the instructions, datapaths for computing the gate transformation on probability amplitude interacting couples, a selection unit implementing the butterfly selection mechanism, a reordering unit for saving the results in the correct locations of the registers file, a trigonometric unit, for obtaining sine and cosine, and a control unit for the architecture management. \\
Considering the software models' results, discussed in Section \ref{subsec:merit_figs}, a  \textbf{20-bit fixed-point} number representation (2 bits for decimal and 18 for fractional parts) with a \textbf{nearest} approximation mechanism is chosen, reducing the area and complexity of arithmetic operators with respect to the floating point one without compromising significantly the quality of the results. Moreover, this reduces the memory required for storing the probability amplitudes, guiding to more emulable qubits for the same considered platform. \\
In order to facilitate the parallelization of computation, a single instruction is considered for executing each gate on $2^{N_\textrm{qubits}-1}$ couples of data, selected through the butterfly mechanism, known the position of the target and that of the eventual control. This allows the evaluation couple-by-couple of the probability amplitudes ---i.e., serially ---, all in parallel --- through a  Single Instruction Multiple Data (SIMD) approach ---, or balancing serial and parallel execution. Parallel evaluations can be obtained by exploiting the absence of data dependencies between interacting couples for a single-gate execution. Identical processing modules are replicated to operate concurrently on different data streams, selected according to a butterfly-based routing mechanism. These modules can either be managed collectively through a centralized control unit or individually through local control units synchronized by shared global control signals.
The windowing mechanism can be harnessed to implement a partial parallelization. It's worth noting that both full serial and full parallel architectures can be viewed as special cases of this mechanism. \\
The full parallel architecture developed as the initial focus of this project, and the windowed one are detailed in the following.

\subsubsection{Full parallel}
\begin{figure*}[t]
	\centering	\includegraphics[width=0.9\textwidth]{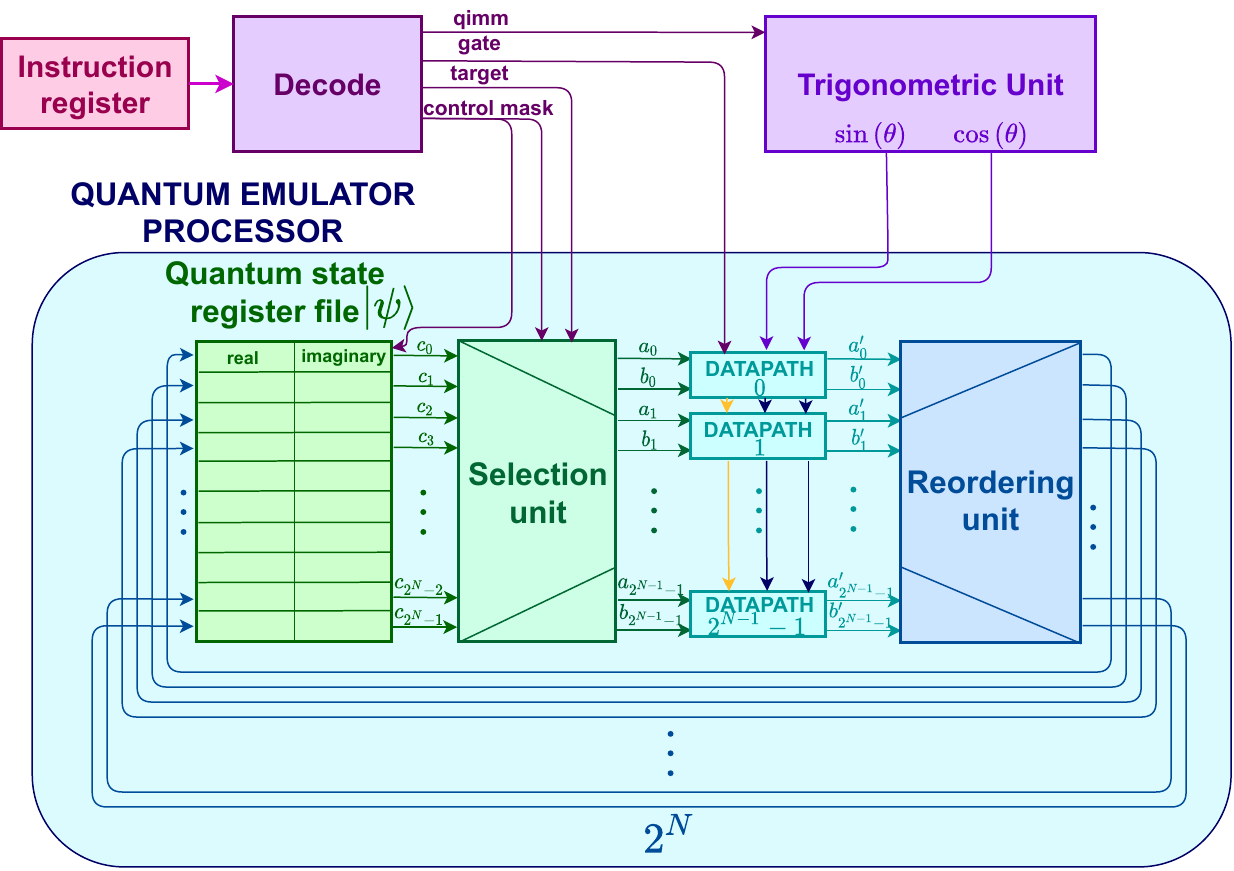}
\caption{Architecture of the full parallel quantum emulator processor. }
\label{fig:AEQUAM_parallel}
\end{figure*}
\begin{figure*}[t]
	\centering	\includegraphics[width=0.9\textwidth]{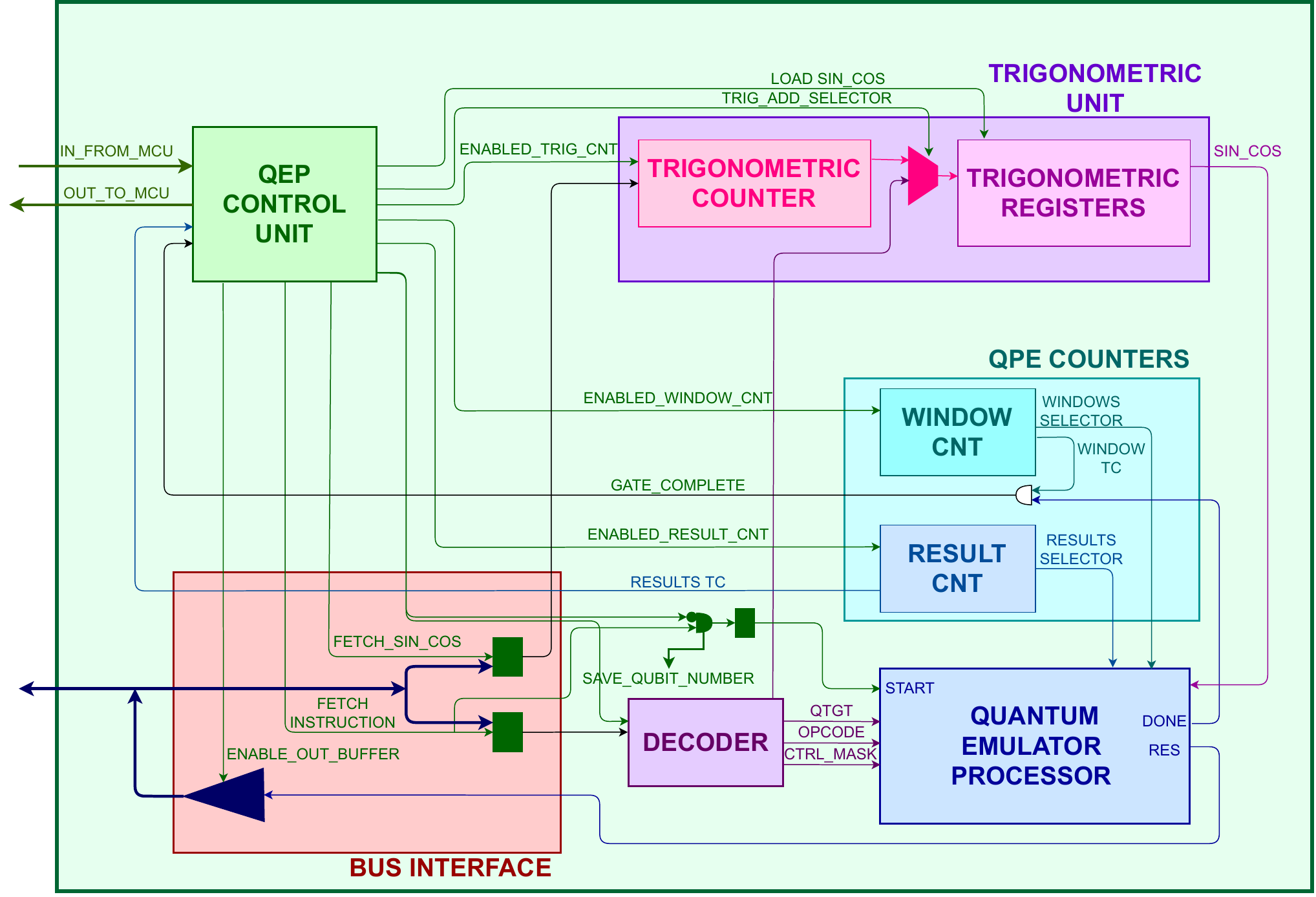}
\caption{Top view emulator}
\label{fig:TopViewEmulator}
\end{figure*}

Figure \ref{fig:AEQUAM_parallel} shows the architecture of the full-parallel quantum emulator processor (QPE). As can be observed, it comprises an instruction register, a decoder, a trigonometric unit,  a quantum state register file, a selection unit, a reordering unit, and $2^{N_{\textrm{qubits}}-1}$ datapaths, one for each interacting couple. The QPE integrates into the more complex architecture shown in Figure \ref{fig:TopViewEmulator}, including a control unit, a set of counters, and a bus interface.  
\\
The instruction register stores the current instruction, which the decoder decomposes into constituent parts, i.e., the target qubit, the control mask, the gate opcode, and immediate (qimm).    \\
The trigonometric unit consists of a register file where the sine and cosine values, computed during compilation and utilized in the quantum circuit, are stored before circuit emulation during the initialization stage. These values are populated using a dedicated counter (trigonometric counter),  generating addresses indicating whether the stored number is sine or cosine.   Following initialization, the immediate instruction field serves as the register file address for selecting both sine and cosine necessary for executing rotational gates. This naive solution offers low computational delay and accurate trigonometrical values. This may be limiting for variational circuits, where a high amount of different trigonometrical values is expected to be used. In the future, other solutions based on the development of efficient dedicated hardware for sine and cosine computation will be investigated to ensure higher flexibility and more limited memory requirements.\\
The quantum state register file has dimensions $2^N_q \times N_{\textrm{bits}} \times 2$, where $N_q$ is the number of qubits, $N_{\textrm{bits}}$ is the number of bits chosen for number representation. It stores the real and imaginary parts of each probability amplitude in each row. For the full-parallel architecture, simultaneous access to all rows is required hence $2^N$ input and output ports are available. \\
The selection and reordering units contain $N$-way multiplexers, with the target qubit as the selector. The selection unit implements the mathematical principle discussed in Section  \ref{sec:butterflyMechanism} for selecting the interacting couples for each datapath. On the other hand, the reordering unit complements this operation by connecting each datapath output to the appropriate state register.
\begin{figure}[t]
	\centering	\includegraphics[width=1\columnwidth]{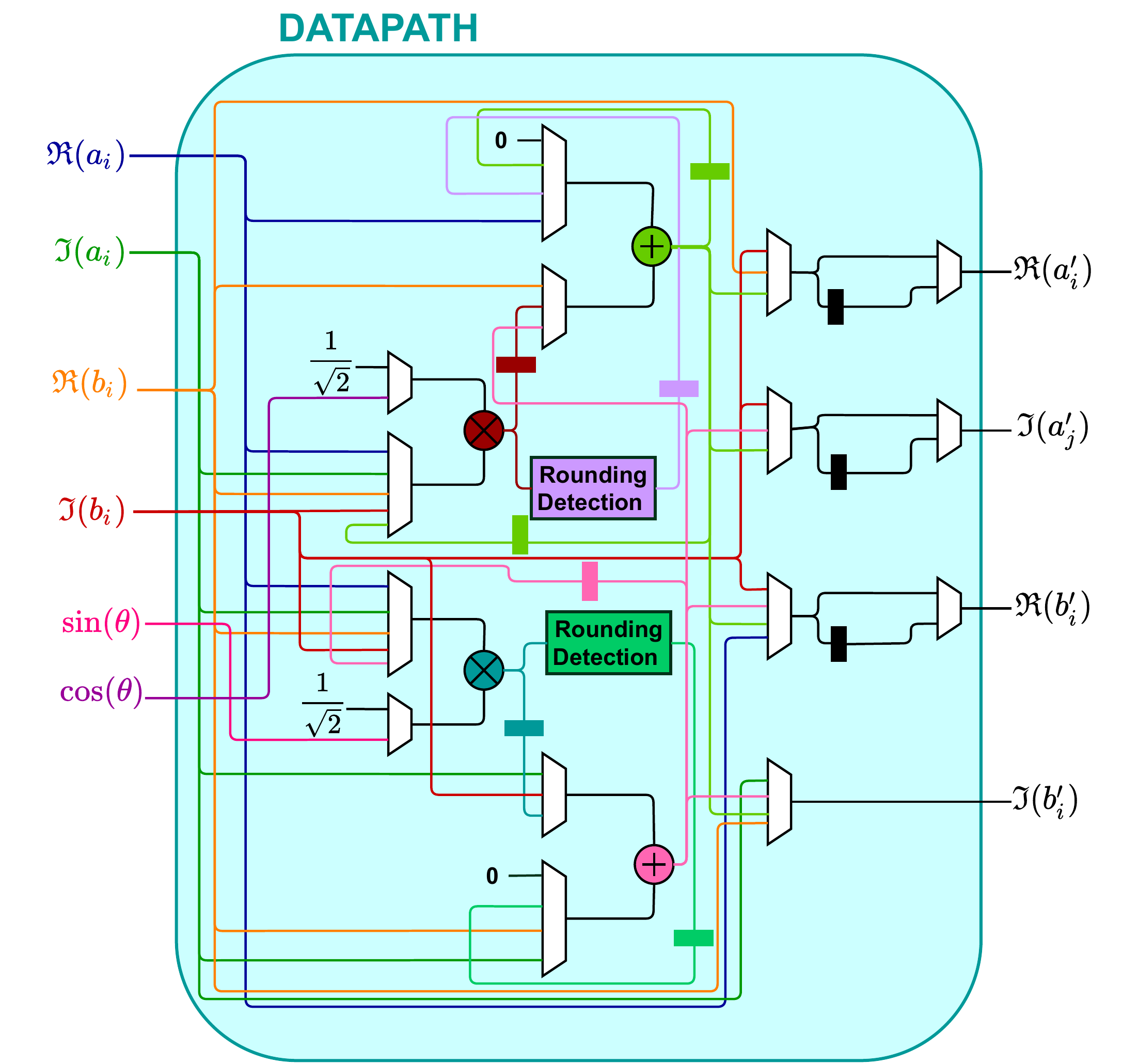}
\caption{Quantum emulator processor datapath. }
\label{fig:DatapathAEQUAM}
\end{figure}\\
The architecture \textbf{datapath}, shown in Figure \ref{fig:DatapathAEQUAM}, utilizes resource sharing to minimize the required area, employing two adders and two multipliers with data dependencies inside the datapath. Arithmetic operators are currently implemented using behavioral descriptions, with the final implementation choice left to the synthesizer.
Operations performed by the datapath and required clock cycles are \textbf{gate-dependent} and determined by the datapath's control unit. 
The control unit employs a \textbf{u-Read-Only-Memories} (\textbf{u-ROMs}) approach, dividing into two u-ROMs for non-rotational and rotational execution. Opcode selection enables the choice of the u-ROM via its MSB, which is also the address of the first state. It is also the address of the first state. Upon activation of the start signal, gate execution commences from the corresponding starting address, progressing through memory locations until reaching the last state and activating the done signal. The SIMD architecture permits control unit sharing among datapaths, enabling VHDL generation to synthesize a control unit for each arithmetic unit or a single one. However, we recommend synthesizing a control unit for each datapath due to lower area requirements than connecting individual control units. \\
The results counter selects the probability amplitude for output at execution's conclusion, while the window counter is not utilized in this architecture. 
The bus interface and the external control unit are addressed in Section \ref{sec:hardware}, as their implementation is contingent on board and communication protocol choices.

\subsubsection{Windowing mechanism}
\begin{figure*}[t]
	\centering	\includegraphics[width=0.9\textwidth]{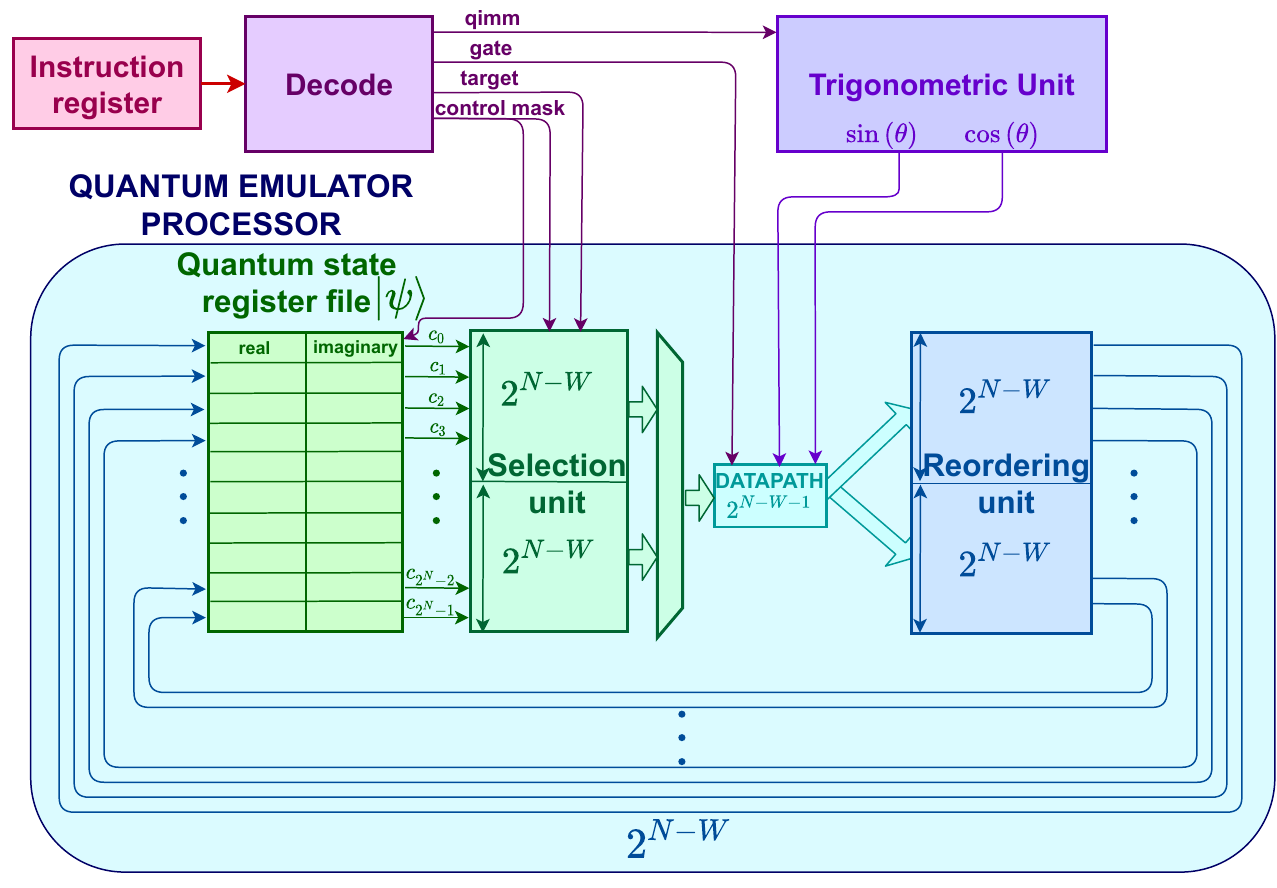}
\caption{AEQUAM windowing architecture. }
\label{fig:AEQUAM_windowing}
\end{figure*}
The windowing architecture shown in Figure \ref{fig:AEQUAM_windowing},   offers a solution for synthesizing a high number of qubits by reducing the level of execution parallelism,  regardless of the execution time. Indeed, while the resources necessary for the quantum state register file and selection/reordering remain fixed, the area required for the datapath can be minimized through resource sharing. This strategy, called windowing, operates by processing distinct windows of probability amplitude couples at a time. Consequently, the total number of datapaths is reduced from $2^{N-1}$ to $2^{N-W-1}$, where $W$ signifies the chosen windowing order and $N$ is the number of qubits. By opting for the highest feasible windowing order, a single datapath suffices for emulation, leading to sequential execution, while selecting a windowing order of 0, it is possible to recover the full-parallel architecture described above. Each additional windowing order doubles the time required to compute a quantum gate but reduces the required amount of area. However, this penalty would be partially compensated in the future by implementing datapaths with enough pipeline stages to enable working in a full-pipe mode since working with different windows removes the data dependency between input and output values, which are present in the fully parallel approach. If the number of pipe levels is equal to the windowing order, the final result would be an almost complete compensation for the delay overhead.\\
The implementation of this architecture leverages the same fundamental block as the full-parallel architecture, with the selection of the operating window performed through the window counter. \\
The main advantage of this approach lies in its flexibility, enabling users to finely select the best architecture for their requirements based on factors such as available board size and quantum circuit dimensions.

\subsection{VHDL generator}\label{sec:VHDLgenerator}
The VHDL description of the most suitable architecture for the user's needs is automatically generated by exploiting Python scripts, which also update the configuration file for the compiler. The scripts are optimized to avoid unnecessary regeneration of already available sub-blocks.\\
The generation process allows users to customize the architecture according to their specific needs. The following degrees of freedom can be adjusted:
\begin{itemize}
\item $N$, which is the number of qubits;
\item $W$, which is the windowing order;
\item $S$, which is the sharing factor for the control units (0 is the recommended value);
\item $Q$, which is the parallelism of the immediate instruction argument.
\end{itemize}
Upon completion, the output of the generation process is the VHDL description corresponding to the architecture shown in Figure \ref{fig:TopViewEmulator}.  

\section{Hardware implementation}\label{sec:hardware}
This section discusses the actual hardware implementation, beginning with the target board choice, followed by the development of software and firmware for interfacing with the board and concluding with an overview of the hardware and communication interfaces.

\begin{figure}[h]
	\centering	\includegraphics[angle=-90,width=0.5\textwidth]{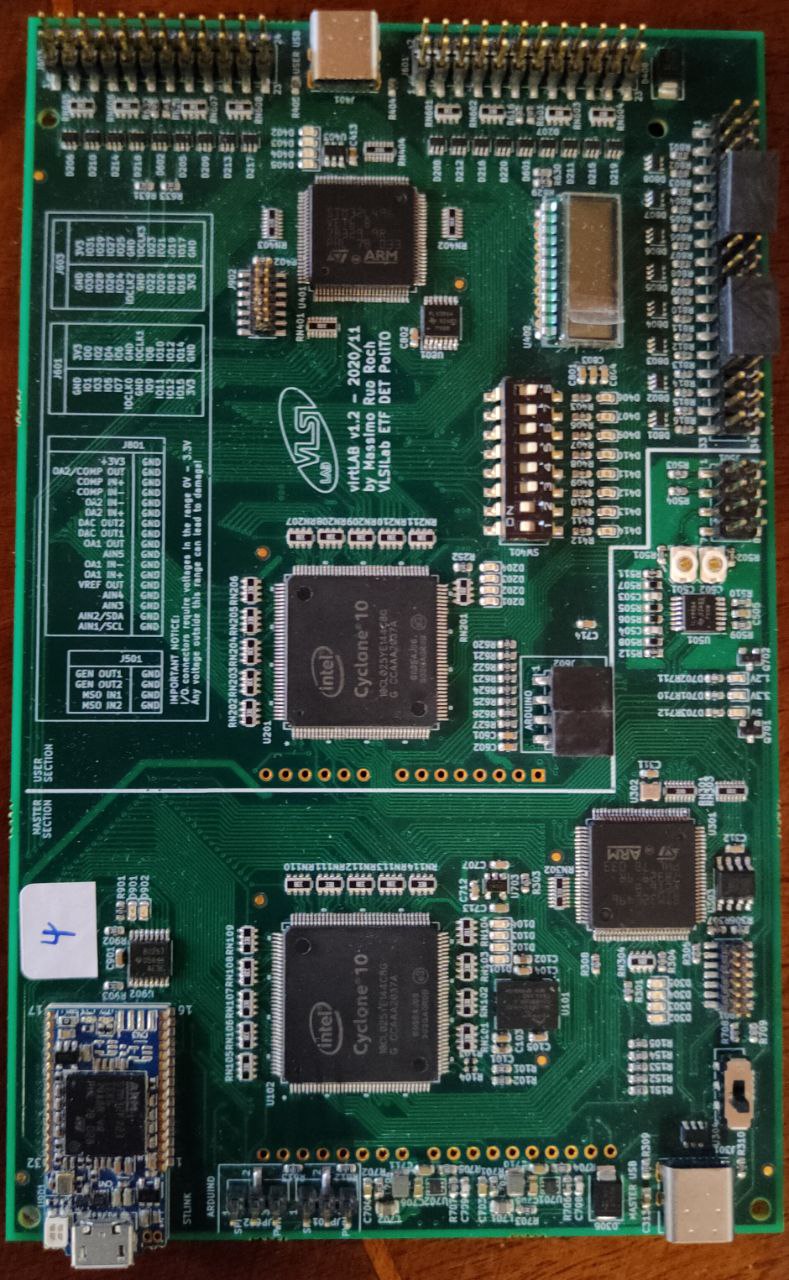}
\caption{The VirtLab board presented in \cite{virtlab}.}
\label{fig:VIRTLAB}
\end{figure}
\subsection{Target board}
For the actual implementation of the proposed architecture, we have selected the VirtLAB board \cite{virtlab}, shown in Figure \ref{fig:VIRTLAB}, which is equipped with two \textbf{Intel Cyclone 10LP 10CL025YE144C8 FPGAs} and two S\textbf{TM32L496 microcontrollers} (\textbf{MCUs}). The board is divided into a user side and a master side; each side hosts one FPGA and one MCU. The master side is particularly useful for debugging, featuring portable benchtop equipment capabilities, including a digital oscilloscope that is accessible through a Java-based Graphic User Interface (GUI). It is a board proposed for teaching purposes since it is cheap and, thanks to the master side, it can partially substitute the benchtop equipment for the applications debugging. Although it has limited area availability, this FPGA was chosen for this work because the emulator is intended to be used in a Master of Science course to verify quantum algorithms, necessitating a low-cost FPGA. Additionally, the current implementation serves as a prototype to evaluate the approach's effectiveness. Indeed, this hardware implementation can be considered a \textbf{proof of concept} for the emulator architecture that, properly adapting the interface, can be synthesized on other bigger and more performant FPGAs. 

\subsection{Software}
To send RISC-like instructions to the board, the proposed toolchain utilizes a Python script. This script reads the instructions and transmits them character by character via a \textbf{serial connection to the board's MCU}. The instructions are encoded in hexadecimal format, with negative numbers represented in modulus form followed by a minus sign, reducing the character count sent.
In addition to instructions, the user needs to provide other types of information to the board, for which the following communication protocol has been established: 
\begin{itemize} 
\item \texttt{?value\#} - Specifies the number of sine and cosine values. 
\item \texttt{\*value\#} - Indicates the number of qubits in use.
 \item \texttt{<value\#} - Sends a sine or cosine value. 
\item \texttt{>value\#} - Transmits an instruction.
 \item \texttt{!} - Signals the end of emulation. 
\end{itemize}
This protocol uses ASCII characters for data transmission, with defined start and end symbols for each value. This method eliminates the need for transmitting redundant leading zeros, as fixed-length values are not required.
After the emulation process, the launcher awaits the retrieval of probability amplitudes from the emulating system, delivered in the format:
\begin{itemize}
 \item \texttt{value\textbackslash n} - Represents a real or imaginary part of a probability amplitude. 
\end{itemize}
Each pair of received values (real and imaginary parts) is then stored in a file.

\subsection{Firmware}
The MCU plays the role of a bridge between the user PC and the FPGA, receiving and propagating instructions and probability amplitude, requiring the development of a proper firmware. In particular, it has two different phases:
\begin{itemize}
\item \textbf{Writing phase}, in which waits to receive characters from the USB serial connection. Once they arrive, the MCU collects the PC’s message. Finally, it writes the desired value/instruction on the bus shared with the FPGA following the handshake protocol.
\item \textbf{Reading phase}, in which the MCU samples the value on the shared
bus according to the handshake protocol and returns it to the user through the USB connection.
\end{itemize}

\subsection{Communication interface} \label{sec:communicationInterface}
\begin{figure}[h]
	\centering	\includegraphics[width=0.5\textwidth]{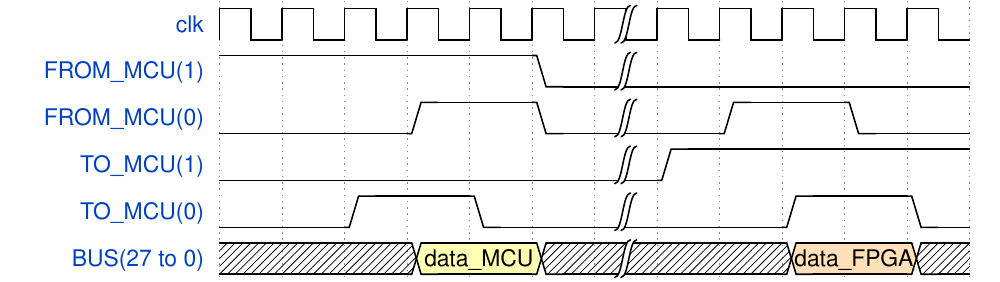}
\caption{Double handshake protocol. }
\label{fig:Handshake}
\end{figure}

On the VirtLAB board, the User MCU and FPGA are directly connected through a 32-bit bus. A \textbf{double handshake protocol} has been considered to ensure robust synchronization between the MCU and FPGA (Figure \ref{fig:Handshake}).  This protocol relies on two main two-bit signals \texttt{FROM\_MCU}, set by the MCU, and \texttt{TO\_MCU},  set by the FPGA. In this implementation, the protocol utilizes the Most Significant Bits (MSBs) of the two signals to indicate changes in the emulation phase, such as role reversals between the MCU and FPGA as receiver or transmitter. The Least Significant Bits (LSBs) are used for the actual handshake protocol.
\begin{figure*}[t]
	\begin{subfigure}[t]{1\columnwidth}
	    \centering
	    \includegraphics[width=\textwidth]{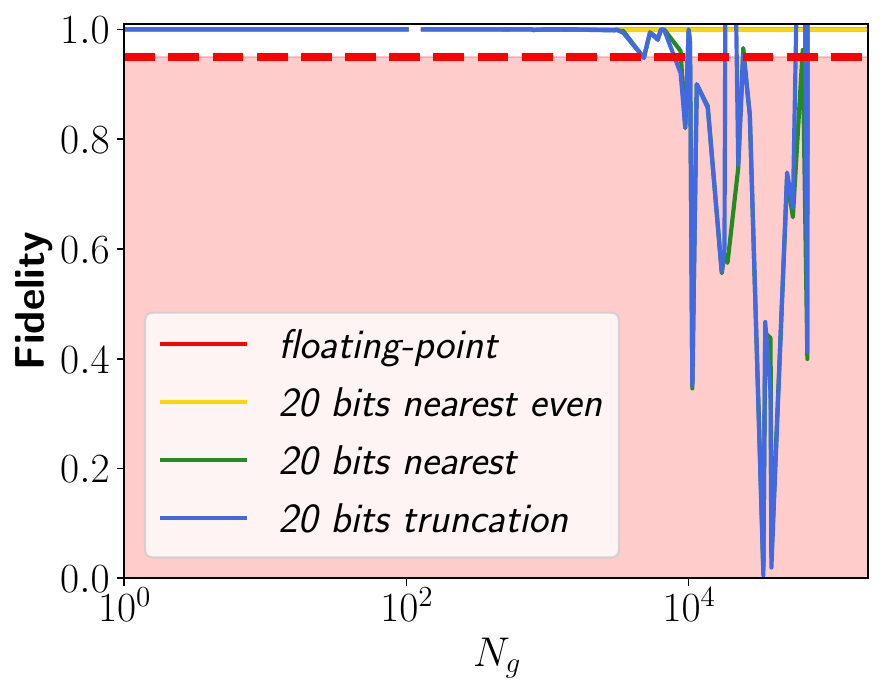}
	    \caption{Helinger Fidelity}
	    \label{fig:HF}
	\end{subfigure}
       \quad\quad
 	\begin{subfigure}[t]{1\columnwidth}
	    \centering
\includegraphics[width=1\textwidth]{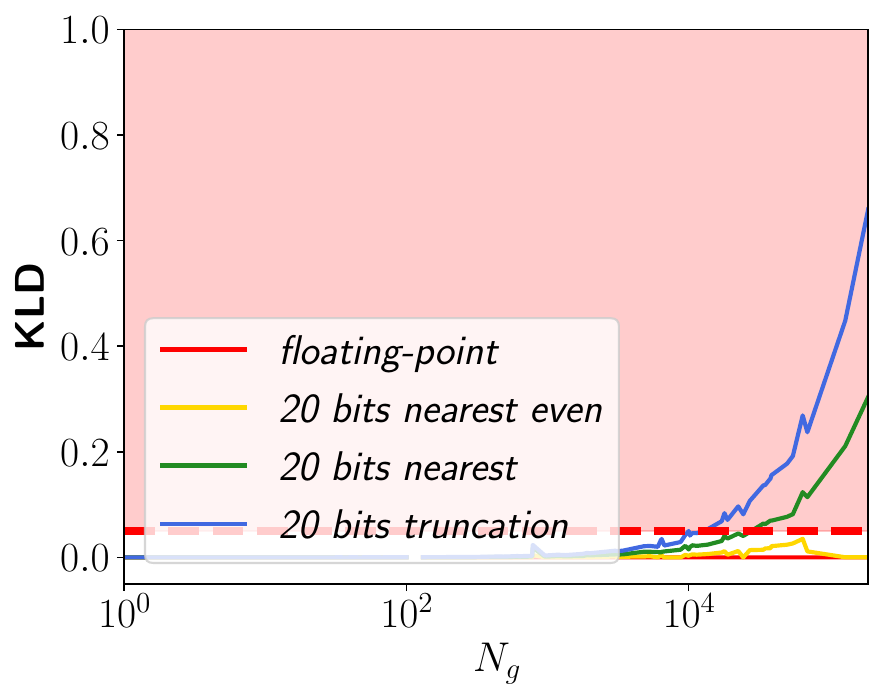}
	    \caption{Kullback Leibler Divergence (KLD)}
	    \label{fig:KLD}
	\end{subfigure}
         
 	\begin{subfigure}[t]{1\columnwidth}
	    \centering
\includegraphics[width=1\textwidth]{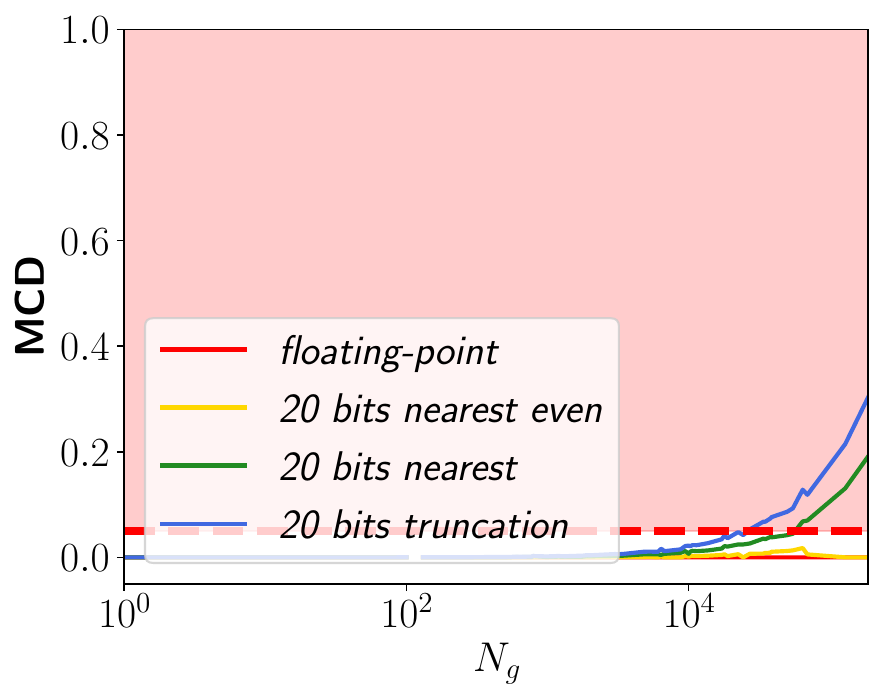}
	    \caption{Maximum Complex Distance (MCD)}
	    \label{fig:MCD}
	\end{subfigure}
 \quad\quad
  	\begin{subfigure}[t]{1\columnwidth}
	    \centering
\includegraphics[width=1\textwidth]{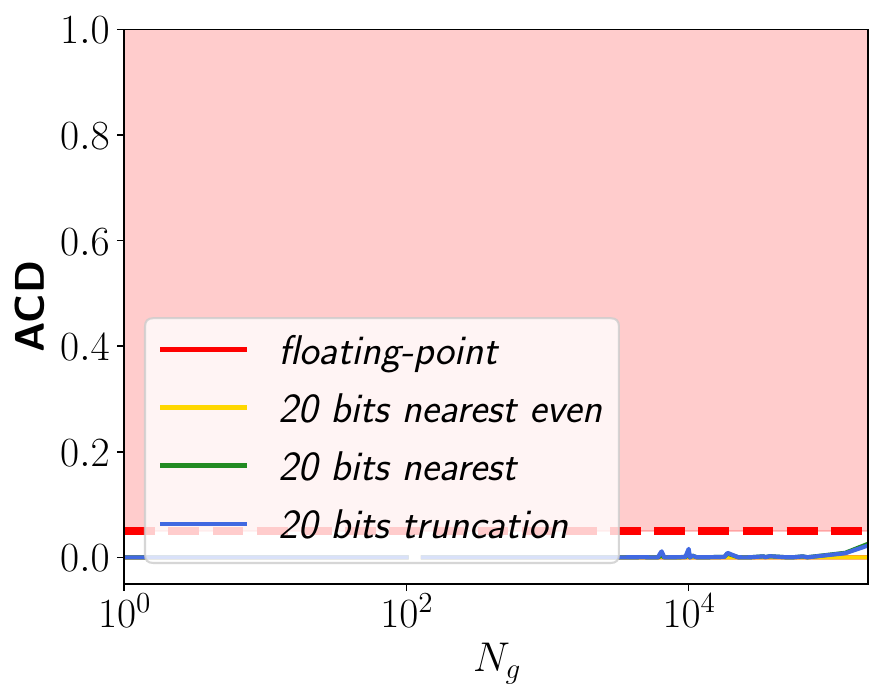}
	    \caption{Average Complex Distance (ACD)}
	    \label{fig:ACD}
	\end{subfigure}
	\caption{Figures of merit to estimate the emulation quality as a function of the number of gates involved in the circuit ($N_g$) varying the number representation, considering benchmarks of the publicly accessible repositories.}
	\label{fig:Results}
\end{figure*}
\subsection{Hardware}
The architecture shown in Figure \ref{fig:TopViewEmulator} has been synthesized on the user FPGA (Intel Cyclone 10LP 10CL025YE144C8G) of the target board using Quartus, employing a twenty-bit data parallelism. Various configurations, including alterations in the number of qubits and windowing order degrees of freedom, were explored to evaluate its potential and delineate operational limits.\\
To ensure seamless communication and mitigate conflicts, the input/output bus connecting the MCU and FPGA incorporates a tri-state buffer controlled by an external control unit, called the Quantum Emulator Processor (QEP) Control Unit. \\
To streamline the design and isolate unrelated segments within the top entity, two distinct fetching registers are utilized. One is dedicated to sampling sine and cosine function values, while the other records incoming instructions and the count of utilized qubits.
The QPE Control Unit allows communication with the MCU, implementing the chosen double handshake protocol setting fetch and enable signals.

\section{Results}\label{sec:results}
This section presents relevant software, simulation, and synthesis results. The codes and benchmark circuits employed to obtain these are accessible in the \mbox{\href{https://drive.google.com/drive/folders/1sbn9oUAO3Rgr_xAzV4Cf5xwtV8WIm61u?usp=sharing}{GitHub repository}}.
The benchmarks comprise a combination of publicly accessible repositories and those generated using the mqt-bench tool \cite{quetschlich2023mqt}. The outcomes from both sources are delineated separately to assess the tool's efficacy in offering comprehensive benchmark coverage and validation.
\begin{figure*}[t]
	\begin{subfigure}[t]{1\columnwidth}
	    \centering
	    \includegraphics[width=\textwidth]{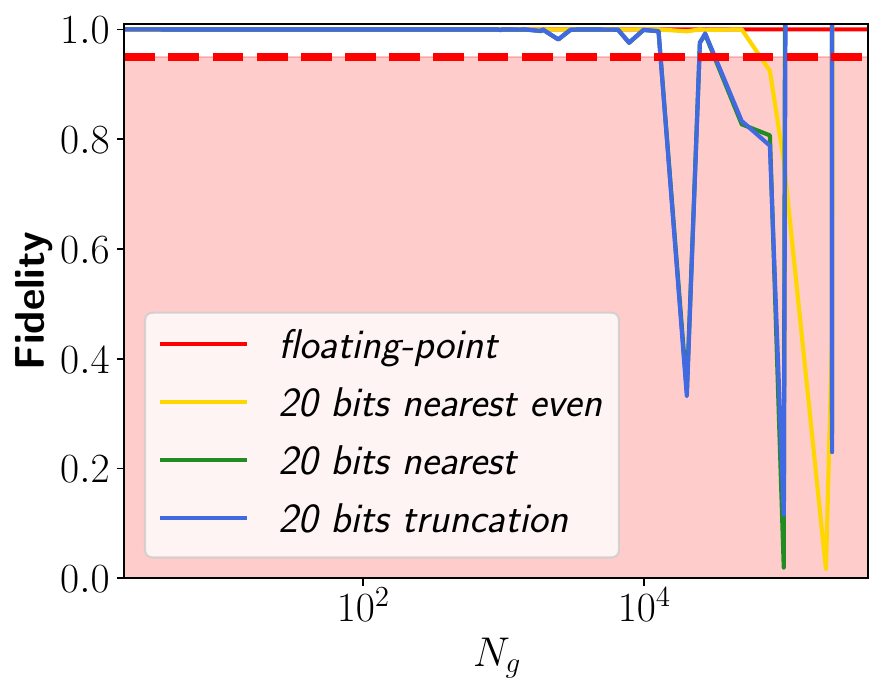}
	    \caption{Helinger Fidelity}
	    \label{fig:HF_mqt}
	\end{subfigure}
       \quad\quad
 	\begin{subfigure}[t]{1\columnwidth}
	    \centering
\includegraphics[width=1\textwidth]{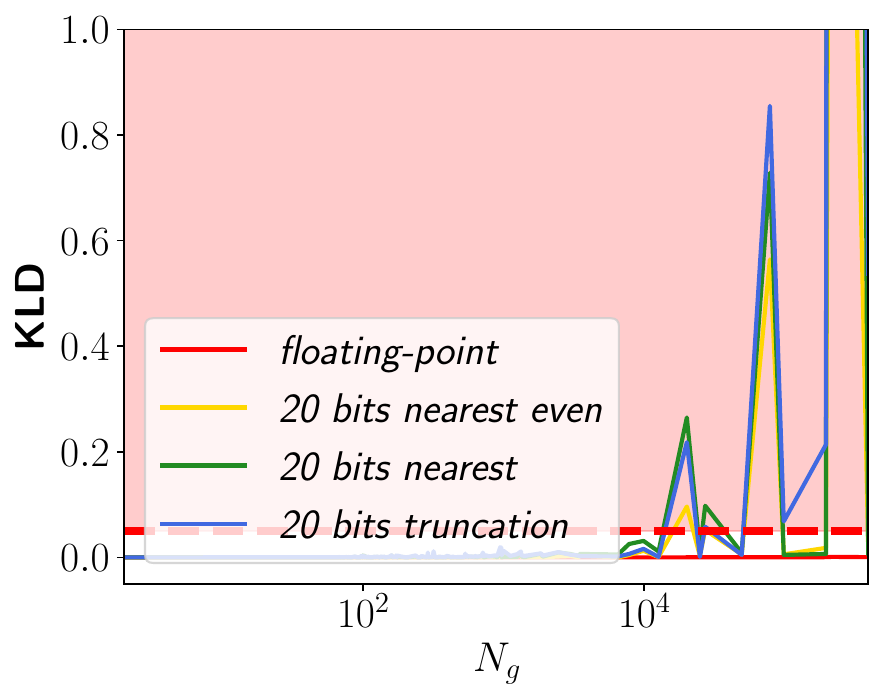}
	    \caption{Kullback Leibler Divergence (KLD)}
	    \label{fig:KLD_mqt}
	\end{subfigure}
         
 	\begin{subfigure}[t]{1\columnwidth}
	    \centering
\includegraphics[width=1\textwidth]{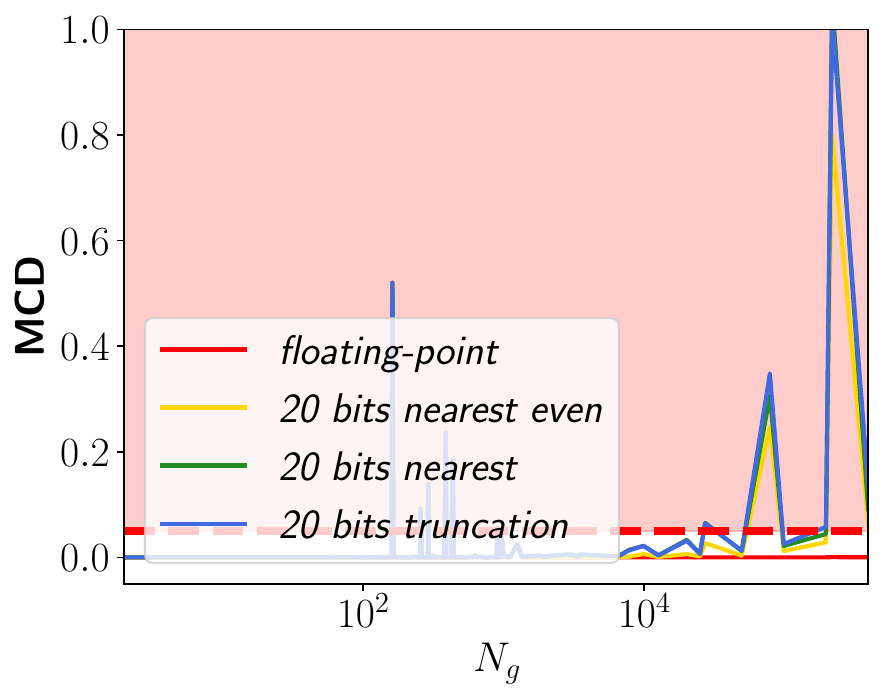}
	    \caption{Maximum Complex Distance (MCD)}
	    \label{fig:MCD_mqt}
	\end{subfigure}
 \quad\quad
  	\begin{subfigure}[t]{1\columnwidth}
	    \centering
\includegraphics[width=1\textwidth]{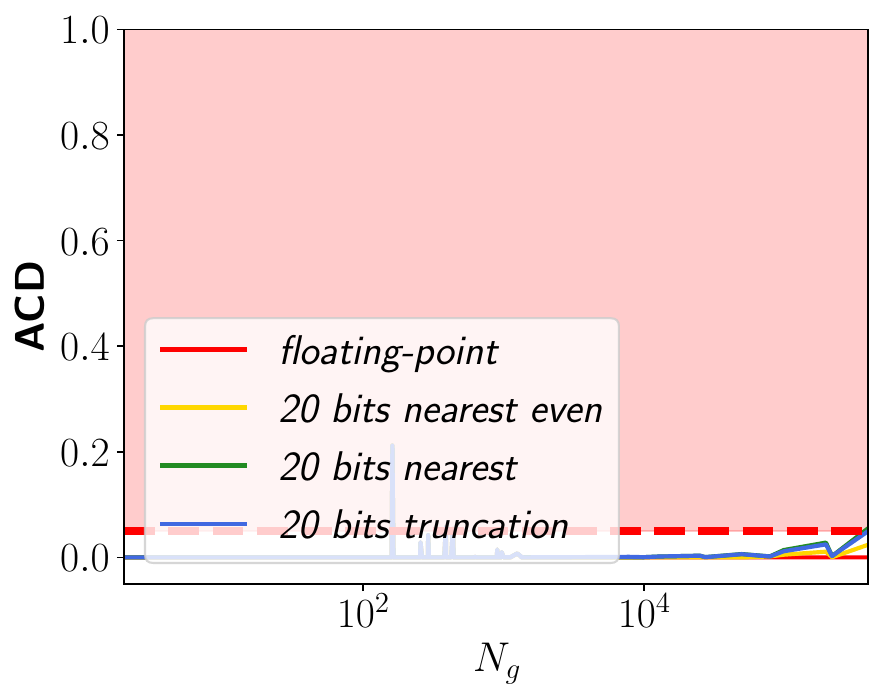}
	    \caption{Average Complex Distance (ACD)}
	    \label{fig:ACD_mqt}
	\end{subfigure}
	\caption{Figures of merit to estimate the emulation quality as a function of the number of gates involved in the circuit ($N_g$) varying the number representation, considering benchmarks of the mqt-bench tool \cite{quetschlich2023mqt}.}
	\label{fig:Results_mqt}
\end{figure*}
\begin{figure*}[t]
	\begin{subfigure}[t]{\columnwidth}
	\centering
	\includegraphics[width=1\textwidth]{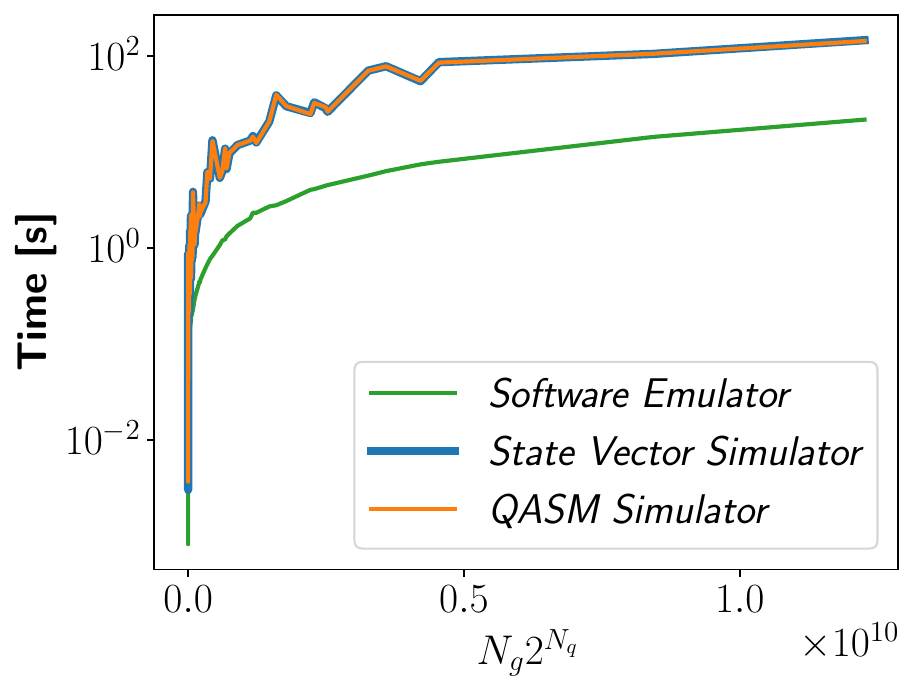}
\caption{Benchmarks of the publicly accessible repositories}
\label{fig:ComparisonTime}
	\end{subfigure}
       \quad\quad
 	\begin{subfigure}[t]{\columnwidth}
	\centering
	\includegraphics[width=1\textwidth]{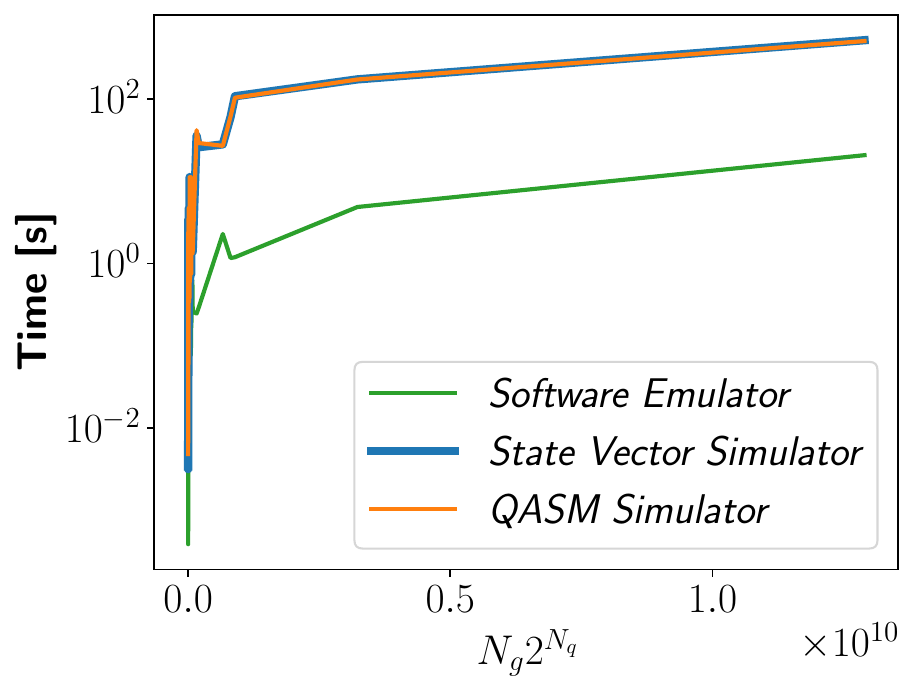}
\caption{Benchmarks of the mqt-bench tool.}
\label{fig:ComparisonTime_mqt}
	\end{subfigure}
 \caption{Comparison of the emulation time required with the proposed approach implemented in software, state vector simulator and QASM simulator available in Qiskit as a function of the circuit complexity evaluated as $N_g 2^{N_q}$, where $N_g$ is the number of gates and $N_q$ the number of qubits, implying $2^{N_q}$ equal to the state vector length.}
  \label{fig:time}
\end{figure*}
\subsection{Sofware results}
This section reports the results obtained with software models presented in Section \ref{sec:SoftwareModels}.
\subsubsection{Setup}
Tests have been conducted on a single-process Intel(R) Xeon(R) Gold 6134 CPU @ 3.20 GHz opta-core, Model 85, with a memory of about 103 GB \cite{processorservervlsi}, comparing our software models with Qiskit state vector and QASM simulators. In the case of fixed-point models, the number of bits considered for number representation has varied from 8 to 32. The benchmark circuits ranged from 2 to 16 qubits.
\subsubsection{Figures of merit}\label{subsec:merit_figs}
To estimate the emulation quality, \textbf{Hellinger fidelity} \cite{noauthor_qiskitquantum_infohellinger_fidelity_nodate} ($[0,1]$) and \textbf{ Kullback Leibler Divergence} (\textbf{KLD}) \cite{kullback_information_1951} divergence have been evaluated for each circuit tested. \\
Fidelity is a measure of the \textit{closeness} of two quantum states, assuming value one if they are identical. In particular, it is defined as $(1-H^2)^2$, where $H$ is the \textbf{Hellinger distance}, computed as:  
\begin{center}
    \begin{equation}
        H(I,R)=\frac{1}{\sqrt{2}}\sqrt{\sum_{i=1}^k(\sqrt{I_i}-\sqrt{R_i})^2} \, ,
    \end{equation}
\end{center}
where $R$ and $I$ are the probability distribution, i.e., the square module of the state vector, of the two quantum states, in this case, obtained from our models and Qiskit state vector simulator, which is considered an ideal reference, respectively. \\
On the other hand, KLD is defined as the \textit{difference} between the two quantum states. Therefore, it is equal to zero in the case of two identical states. The KLD is evaluated as:
\begin{equation}
    D_{KL}(I\|R) = \sum_{x\in\chi}I(x)\log{\Big(\frac{I(x)}{R(x)}\Big)} \, , 
\end{equation}
where $R$ and $I$ are the probability distribution, i.e., the square module of the state vector, of the two quantum states, in this case, obtained from our models and Qiskit state vector simulator, which is considered an ideal reference, respectively. \\
 Unfortunately, these figures of merit are not completely satisfactory. Since both fidelity and KLD compare two probability distributions, their evaluation approach assumes that the sum of all the probability coefficients is one. However, arithmetic approximations can lead to a deterioration of the probability sum to values close to one but different from it. This can cause a divergence of fidelity and KLD in their computation. Moreover, they do not consider eventual phase errors. Therefore, we decided to compute also the maximum and the average complex distance, i.e., the maximum and the average, respectively, distance among the probability amplitudes of the two outcomes: 
 \begin{equation}
     |I_i - R_i | \, .
 \end{equation}
 These figures of merit together with the others allow a complete functional evaluation of the emulation approach. 

\subsubsection{Discussion}
\begin{figure}[h]
	\centering	\includegraphics[width=0.5\textwidth]{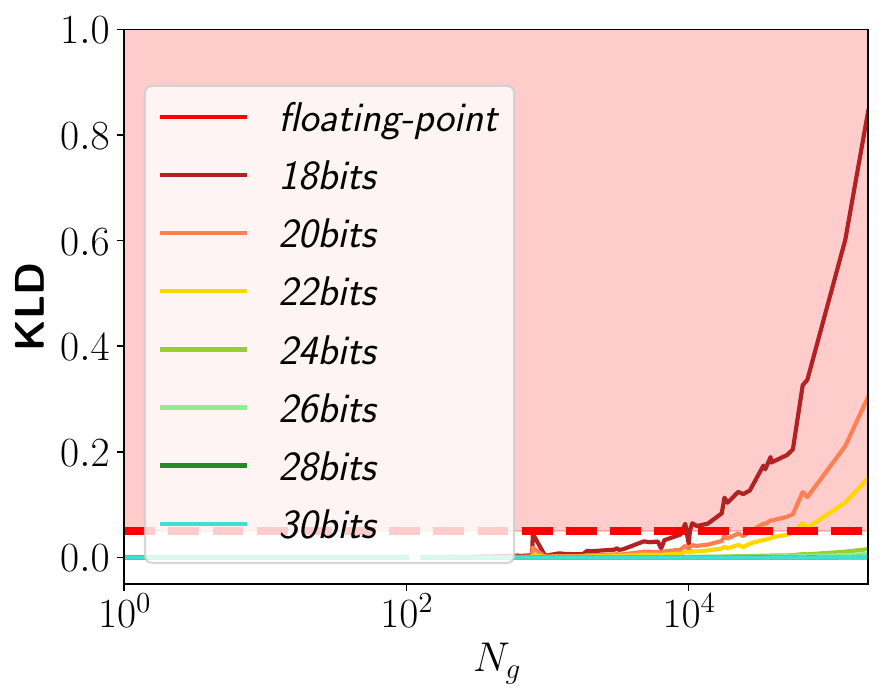}
\caption{ Kullback Leibler Divergence (KLD) as a function of the number of gates involved in the circuit ($N_g$ )
varying the number of bits for nearest number representation, considering benchmarks of the publicly accessible repositories.}
\label{fig:KLD_nearest}
\end{figure}

Figures \ref{fig:Results} and \ref{fig:Results_mqt} illustrate the considered figures of merit for evaluating emulation quality as a function of circuit length, represented by the number of gates ($N_g$). These evaluations include various numerical representations, utilizing benchmarks from publicly accessible repositories as well as those generated using the mqt-bench tool \cite{quetschlich2023mqt}. For fixed-point representation, 20-bit precision is reported as it offers the best compromise between accuracy and complexity, yielding an acceptable arithmetic error with minimal bit usage, as shown in Figure \ref{fig:KLD_nearest}
\\ 
It can be observed that, as expected, fixed-point number representation leads to a reduction in result accuracy, with the impact increasing alongside the circuit length. However, the choice of approximation mechanism is crucial in mitigating arithmetic errors. As mentioned, the truncation mechanism produces the poorest results, while the nearest-even method yields the best outcomes. The nearest rounding method presents a suitable compromise, offering an acceptable quality of results with limited implementation complexity.\\
The peaks observed in the fidelity plots correspond to phenomena previously discussed, specifically the divergence in evaluation when the sum of the probabilities being compared deviates from one, even if only slightly.
Furthermore, the results obtained using mqt-bench have generally a lower quality than those from publicly accessible repositories. This discrepancy arises because the former includes more circuits involving rotational gates, which are subject to greater uncertainty compared to other types of quantum gates due to the increased number of arithmetic operations required.
Based on this analysis, we implemented our architecture with a 20-bit nearest number representation, achieving acceptable accuracy while conserving resources. However, the architecture is flexible enough to accommodate increased precision according to the user's needs.
\begin{figure*}[t]
	\centering	\includegraphics[width=1\textwidth]{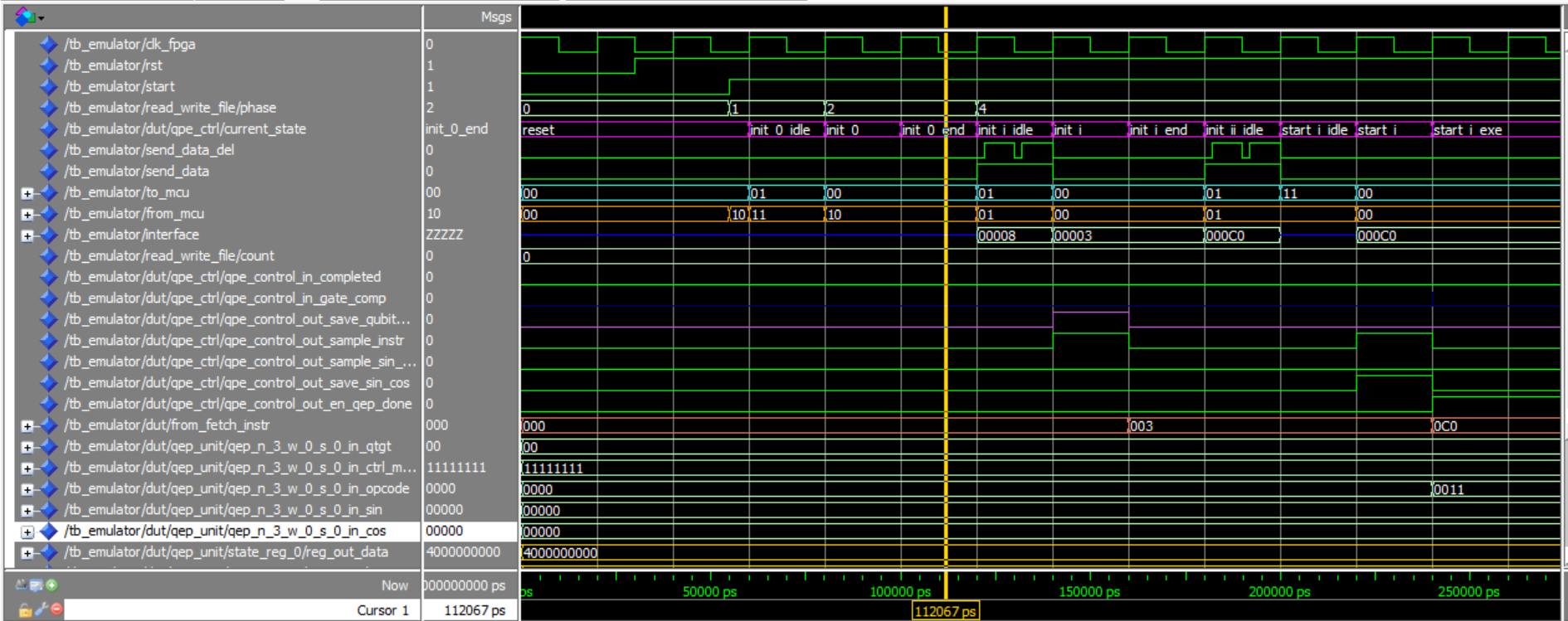}
\caption{Waves of the simulation of a three-qubit bell state circuit emulation. In particular, the emulator initialization is shown. \vspace{-10pt}}
\label{fig:waveBegin}
\end{figure*}
\begin{figure*}
	\centering	\includegraphics[width=1\textwidth]{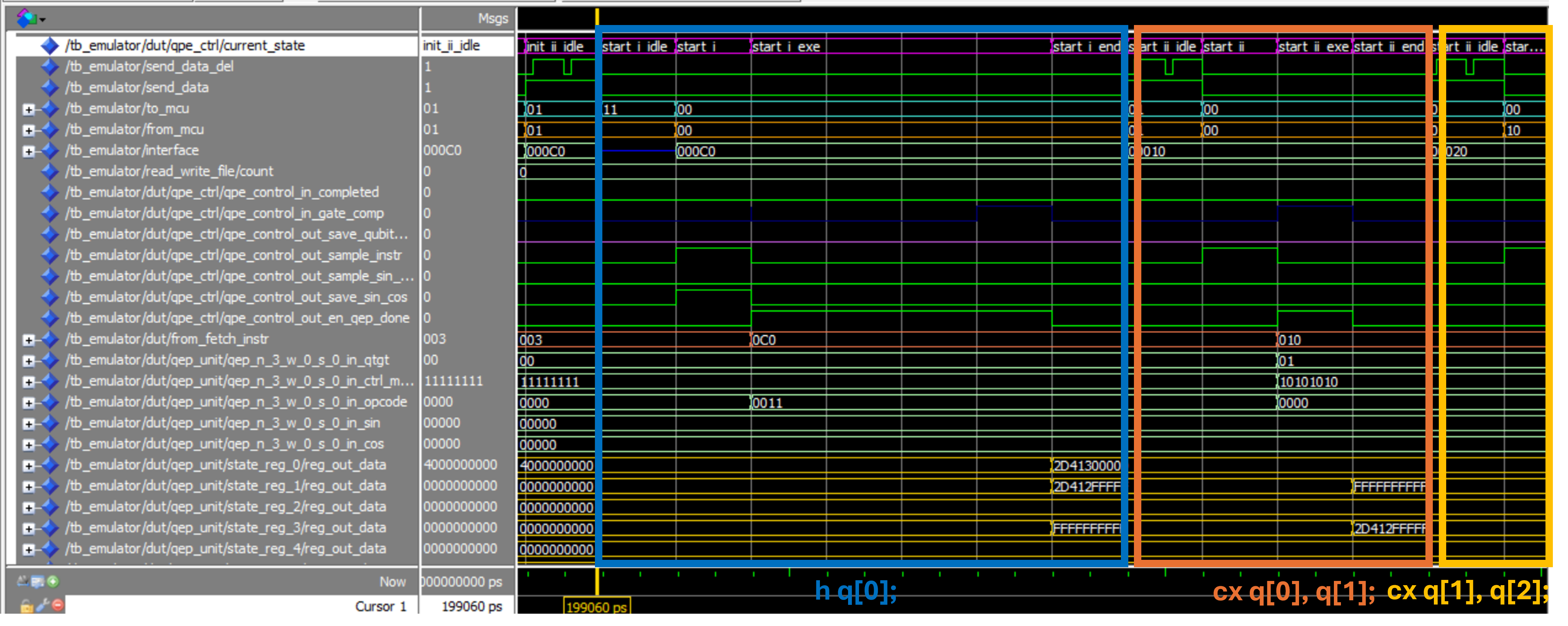}
\caption{Waves of the simulation of a three-qubit bell state circuit emulation. In particular, the gate-by-gate circuit execution is shown. \vspace{-10pt}  }
\label{fig:waveGateExec}
\end{figure*}
\begin{figure*}
	\centering	\includegraphics[width=1\textwidth]{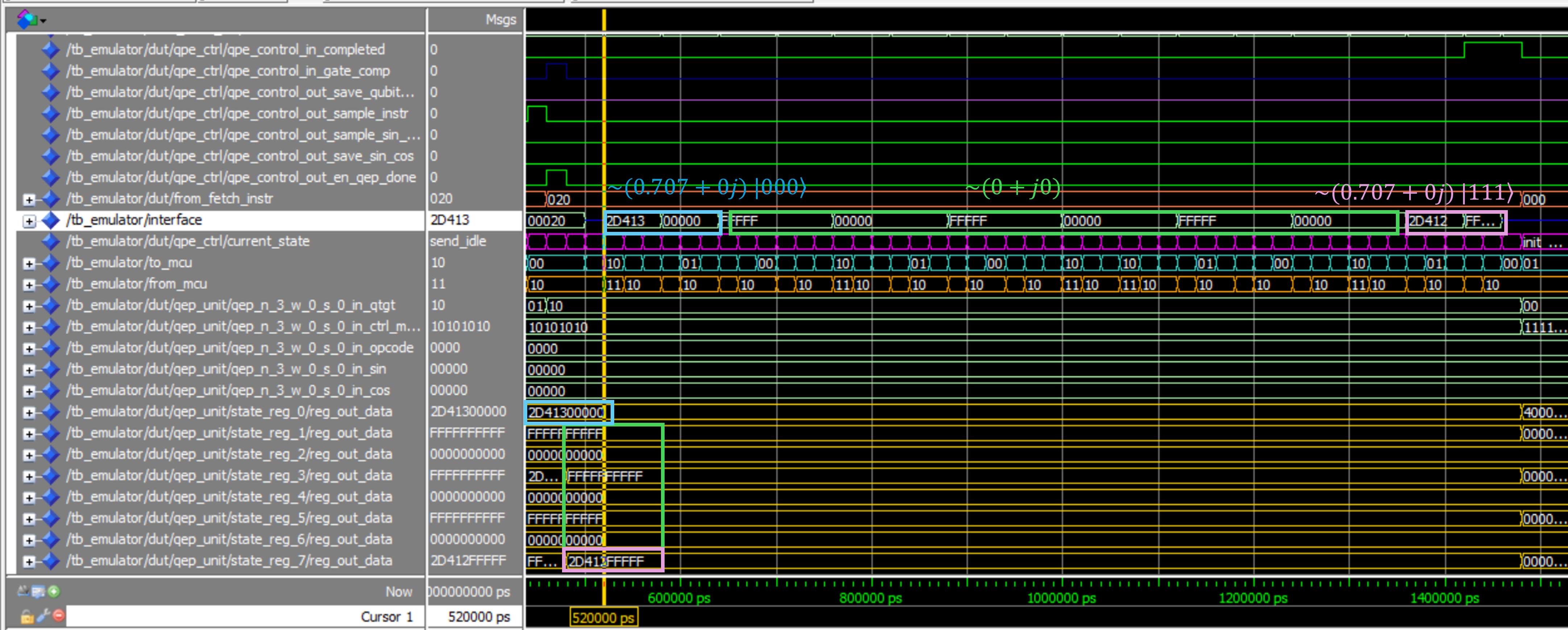}
\caption{Waves of the simulation of a three-qubit bell state circuit emulation. In particular, the final state vector acquisition is shown. \vspace{-10pt}}
\label{fig:waveStateReception}
\end{figure*}\begin{figure*}
	\begin{subfigure}[t]{1\columnwidth}
	    \centering
	    \includegraphics[width=0.92\textwidth]{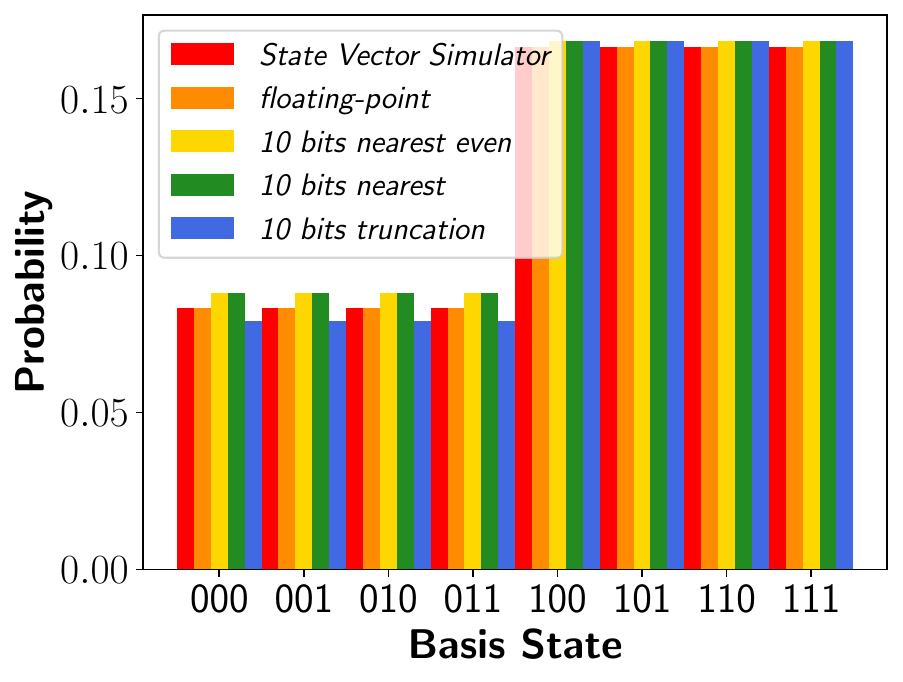}
	    \caption{Probability distribution of the teleport circuit (publicly accessible repositories) executed on the software models and Qiskit state vector simulator }
	    \label{fig:teleport}
	\end{subfigure}
        \quad
 	\begin{subfigure}[t]{1\columnwidth}
	    \centering
\includegraphics[width=0.92\textwidth]{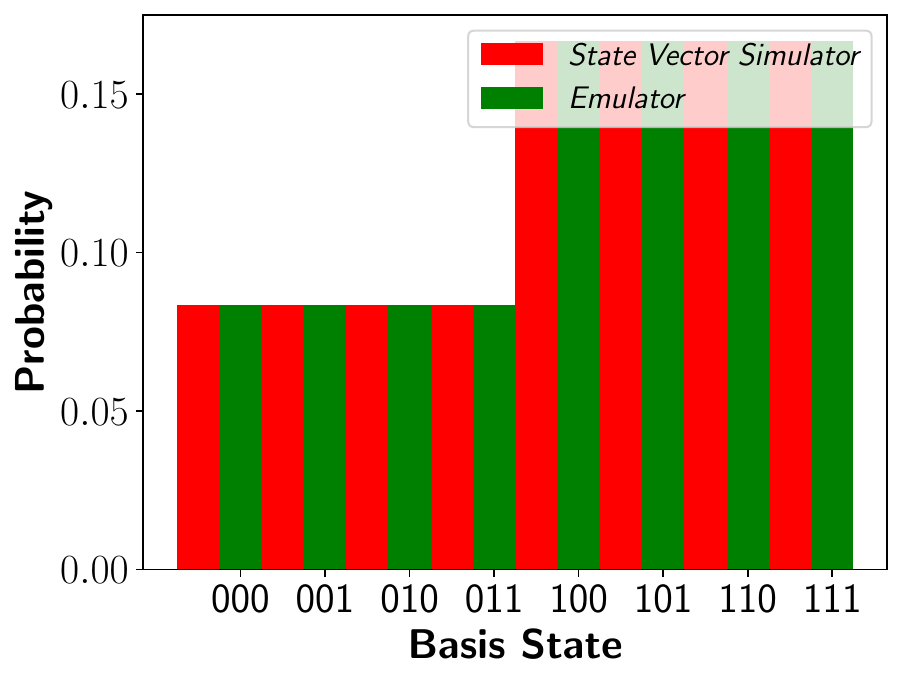}
	    \caption{Probability distribution of the teleport circuit (publicly accessible repositories) executed on the AEQUAM architecture and Qiskit state vector simulator}
	    \label{fig:teleport_vhd}
	\end{subfigure}
	\begin{subfigure}[t]{1\columnwidth}
	    \centering
	    \includegraphics[width=0.92\textwidth]{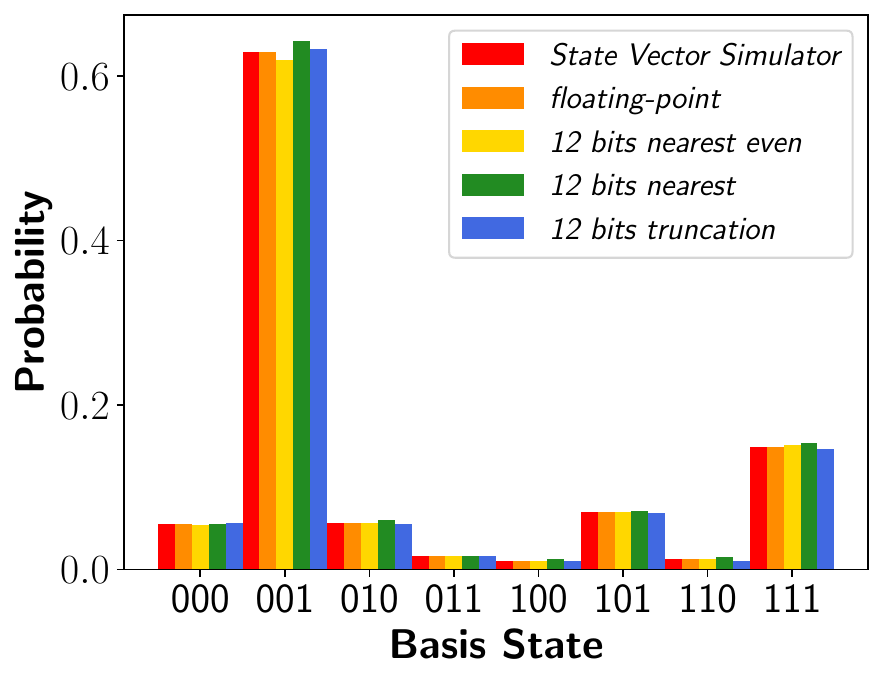}
	    \caption{Probability distribution of a three-qubit Quantum Neural Network (QNN) circuit (mqt-bench tool) executed on the software models and Qiskit state vector simulator }
	    \label{fig:qnn}
	\end{subfigure}
        \quad
 	\begin{subfigure}[t]{1\columnwidth}
	    \centering
\includegraphics[width=0.92\textwidth]{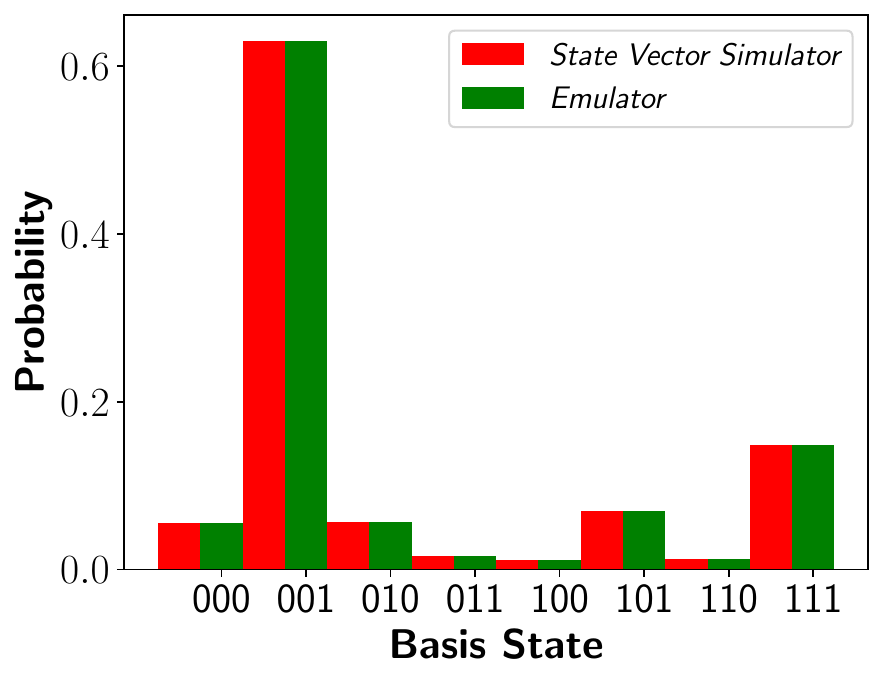}
	    \caption{Probability distribution of  a three-qubit Quantum Neural Network (QNN) circuit (mqt-bench tool) executed on the AEQUAM architecture and Qiskit state vector simulator}
	    \label{fig:qnn_vhd}
	\end{subfigure}
	\caption{Probability distributions of two benchmark circuits. }
	\label{fig:ResultsPlot}
\end{figure*}\\
Finally, Figure \ref{fig:time} shows the emulation time required for the proposed approach implemented in software, the state vector simulator, and the QASM simulator available in Qiskit, as a function of circuit complexity, evaluated as $N_g \cdot 2^{N_q}$, where $N_g$ is the number of gates and $N_q$ is the number of qubits, with $2^{N_q}$ representing the state vector length. It is evident that the proposed approach, even when implemented in software, offers a significant time advantage (at least one order of magnitude) compared to Qiskit's simulators.

\subsection{Functional Verification of the Architecture}
\begin{figure*}[t]
	\begin{subfigure}[t]{\columnwidth}
	\centering
	\includegraphics[width=1\textwidth]{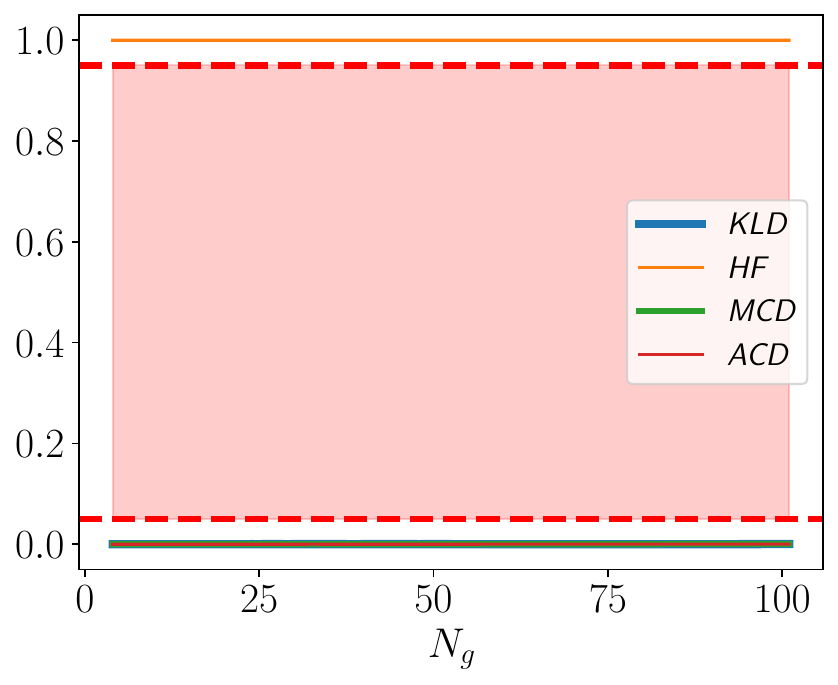}
\caption{Figures of merit to estimate the emulation quality as a function of the number of gates involved in the circuit ($N_g$), considering benchmarks of the publicly accessible repositories.}
\label{fig:vhd_res}
	\end{subfigure}
       \quad\quad
 	\begin{subfigure}[t]{\columnwidth}
	\centering
	\includegraphics[width=1\textwidth]{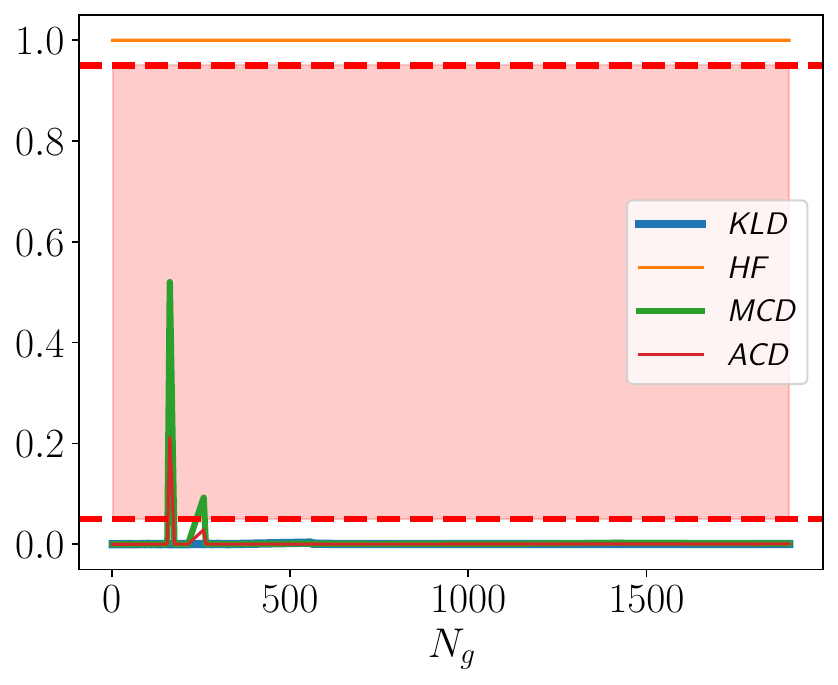}
\caption{Figures of merit to estimate the emulation quality as a function of the number of gates involved in the circuit ($N_g$), considering benchmarks of the mqt-bench tool \cite{quetschlich2023mqt}.}
\label{fig:vhd_mqt_res}
	\end{subfigure}
 \caption{Figures of merit --- Helinger Fidelity (HF), Kullback Leibler Divergence (KLD), Maximum Complex Distance (MCD), and Average Complex Distance (ACD) --- to estimate the emulation quality obtained simulating the benchmark circuits with the proposed architecture. }
  \label{fig:benchmarks_circuits_vhd}
\end{figure*}

The architecture's functional verification was performed through a hierarchical approach. Initially, each sub-block was simulated individually using \textbf{ModelSim} 11.1 tool, leveraging specifically developed testbenches, which are available in the article's GitHub repository. Following the verification of individual components, the complete architecture was tested by setting the emulator size and degree of parallelization, focusing on a selection of representative quantum circuits (involving both Clifford+T and rotational gates).
Subsequently, the robustness of the hardware description was evaluated by automating the simulation of all quantum circuits considered in the software test involving up to eight qubits. This process was managed via a script that \textbf{automates} several tasks: invoking the compiler, adjusting the \textbf{testbench} and \textbf{TCL} (Tool Command Language) scripts necessary for the simulation, and executing the simulation in ModelSim. The simulation results were then compared against the outputs from software models and Qiskit simulators to ensure the accuracy and correctness of the architecture description.
\\
Figures \ref{fig:waveBegin}, \ref{fig:waveGateExec} and \ref{fig:waveStateReception} display the waveforms for executing a three-qubit Bell state circuit on the emulator, employing a fully parallel three-qubit architecture. Specifically, Figure \ref{fig:waveBegin} illustrates the emulator's initialization process, which involves setting the initial state and determining the number of actively used probability amplitudes (i.e., qubits). Additionally, the communication protocol between the emulator and the MCU can be observed as detailed in Section \ref{sec:communicationInterface}.
Figure \ref{fig:waveGateExec}  depicts the step-by-step execution of quantum gates within the circuit and the corresponding evolution of the state vector. Finally, Figure \ref{fig:waveStateReception} shows the transmission of the final state vector to the MCU, where the communication protocol can be again observed. Except for some little variations due to finite-precision arithmetic, the obtained final state is $\frac{1}{\sqrt{2}} (\ket{000} + \ket{111})$ as expected. \\
Figure \ref{fig:ResultsPlot} displays the probability distributions obtained by simulating two benchmark circuits --- one from publicly accessible repositories and another from the mqt-bench --- using software models, the architecture, and the Qiskit state vector simulator. These results demonstrate that \textbf{finite arithmetic precision does not significantly impact the probability distributions} produced by both the software models and the architecture. It is important to note that the software models employed a precision level lower than the 20-bit precision recommended for the architecture and used in hardware tests. This was done to show that, for short and simple quantum circuits, even a reduced number of precision bits can produce accurate probability distributions. While we recommend 20-bit precision as it generally provides satisfactory results, users can select a lower precision based on the specific characteristics of the simulated quantum circuit.
\\
Figure \ref{fig:benchmarks_circuits_vhd}  summarizes the results obtained from the architecture by presenting the figures of merit discussed in Section \ref{subsec:merit_figs} as a function of circuit length, represented by the number of gates ($N_g$). These evaluations cover various quantum circuits involving less than nine qubits, using benchmarks from publicly accessible repositories as well as those generated by the mqt-bench tool \cite{quetschlich2023mqt}.\\
The plots demonstrate that the proposed architecture delivers highly accurate results. Specifically, the Helinger Fidelity and KLD consistently remain close to their ideal values --- one and zero, respectively --- while MCD and ACD are generally below 0.05, a threshold considered to indicate reasonable accuracy.
The two peaks observed in the MCD and ACD metrics correspond to global phase variations, which are not so relevant in the context of a quantum circuit's probability distribution. This is evident as the associated Fidelity and KLD remain at their ideal values. However, these variations are present only in two circuits involving a lot of rotational gates and can be reduced by increasing the precision.\\
To conclude, the emulation quality obtained with the tests on our finite precision architecture with the suggested numbers' precision is, on average, satisfactory. 

\subsection{Synthesis results}
This section reports the results obtained by synthesizing the architecture varying the number of qubits ($N_q$), the windowing order ($W$) and the sine-cosine register file dimension ($2^Q$). 

\subsubsection{Setup}
The synthesis was done by executing the TCL files generated by an automation script using \textbf{Quartus} Prime 17. In order to explore the emulator's scalability, various combinations of the number of qubits ($N_q$), the windowing order ($W$), and the sine-cosine register file dimension ($2^Q$) were considered.

\subsubsection{Figures of merit}
The synthesis results were evaluated based on occupied area, speed, and power consumption. Specifically, the following figures of merit were considered:
\begin{itemize}
    \item \textbf{Logic utilization}, measured in terms of logic elements (\textbf{LE}); 
    \item Required \textbf{registers} (\textbf{REG});
    \item Maximum operating \textbf{frequency}, expressed in \si{\mega \hertz};
    \item \textbf{Dynamic Power} consumption, expressed in \si{\milli \watt} (\textbf{DP}); 
     \item \textbf{Static Power} consumption, expressed in \si{\milli \watt} (\textbf{SP}); 
      \item \textbf{Total Power} consumption, expressed in \si{\milli \watt} (\textbf{Tot P}).
\end{itemize}
These values were obtained from the report files generated by the Quartus synthesis.\\
The most crucial figures of merit for evaluating the architecture are those related to \textbf{area} occupation, i.e., the number of logic elements and registers. The occupied area is critical to determine the FPGA size required to emulate a quantum circuit with a given qubit count.
The power consumption considered in this analysis was estimated using the Quartus power analyzer without applying back-annotation (i.e., without using a transaction file from the simulation). Since power consumption is not the most critical aspect of this design, this estimation, while not perfectly accurate, is considered sufficient.

\begin{figure*}
	\begin{subfigure}[t]{\columnwidth}
	    \centering
	    \includegraphics[width=0.9\textwidth]{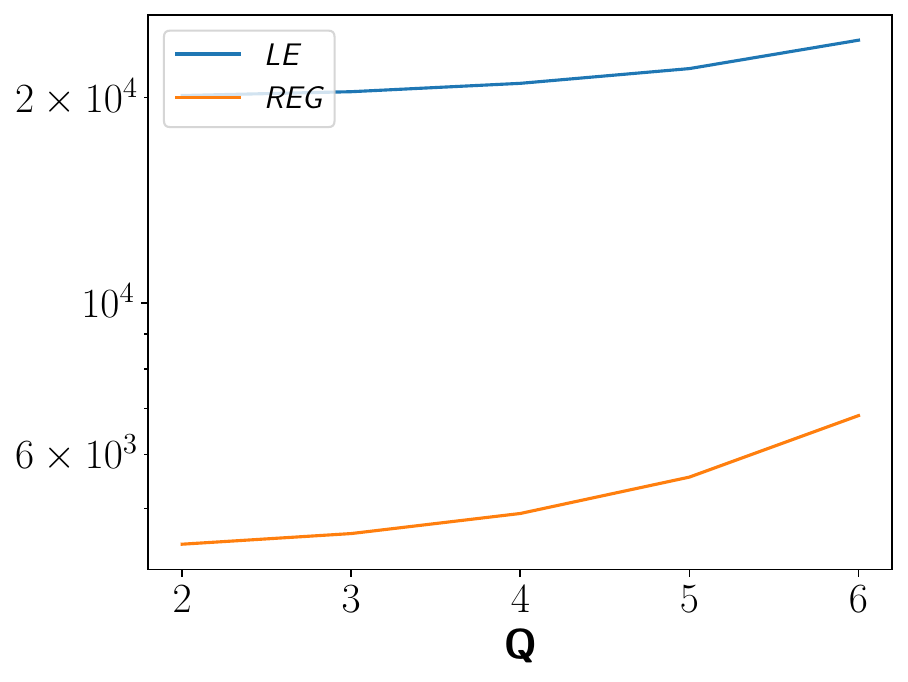}
	    \caption{Number of logic elements (LE) and registers (REG) required as a function of the number of sine-cosine that can be stored ($2^Q$), considering five qubits and the full parallel AEQUAM architecture \vspace{8pt}}
\label{fig:AreaVaryingQ}
	\end{subfigure}
 \quad
 	\begin{subfigure}[t]{\columnwidth}
	    \centering
\includegraphics[width=0.9\textwidth]{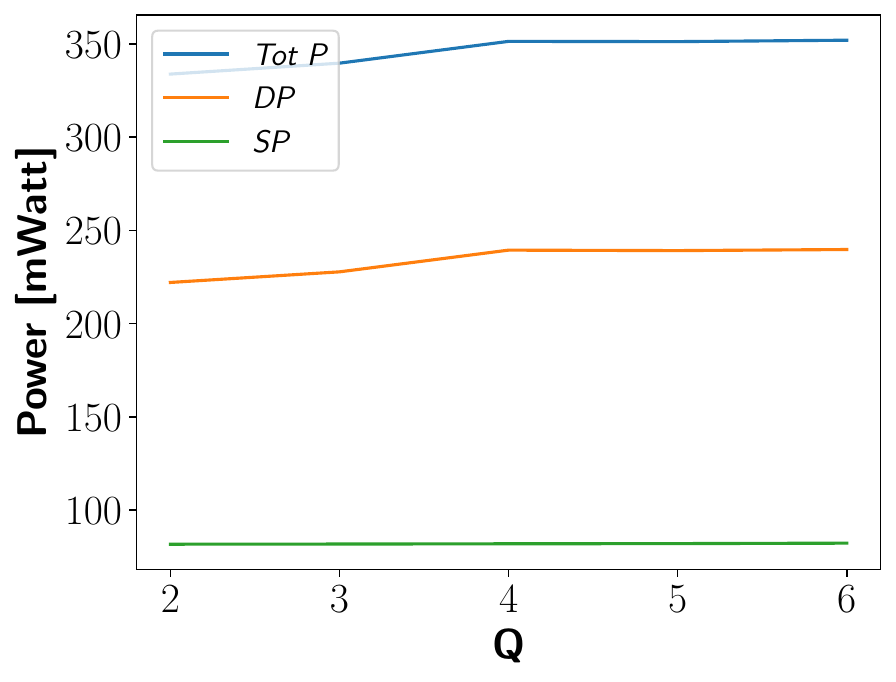}
	    \caption{ Total (Tot P), static (SP) and dynamic power (DP) consumed as a function of the number of sine-cosine that can be stored ($2^Q$), considering five qubits and the full parallel AEQUAM architecture}
\label{fig:PowerVaryingQ}
	\end{subfigure}
 \begin{subfigure}[t]{\columnwidth}
	    \centering
	    \includegraphics[width=0.9\textwidth]{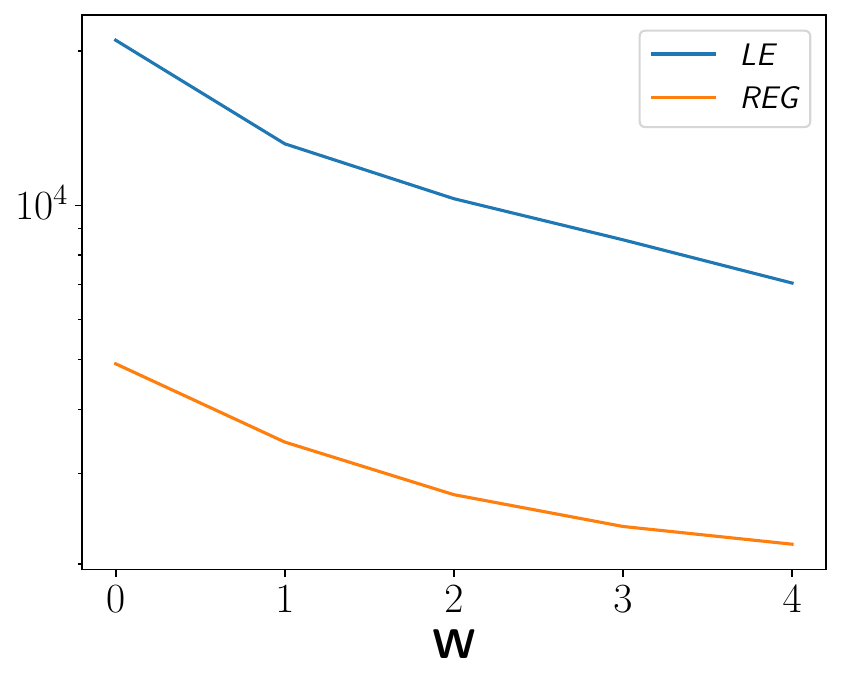}
	    \caption{Number of logic elements (LE) and registers (REG) required as a function of the windowing order ($W$), considering five qubits and Q equal to four \vspace{8pt} }
	    \label{fig:AreaVaryingW}
	\end{subfigure}
 \quad
 	\begin{subfigure}[t]{\columnwidth}
	    \centering
\includegraphics[width=0.95\textwidth]{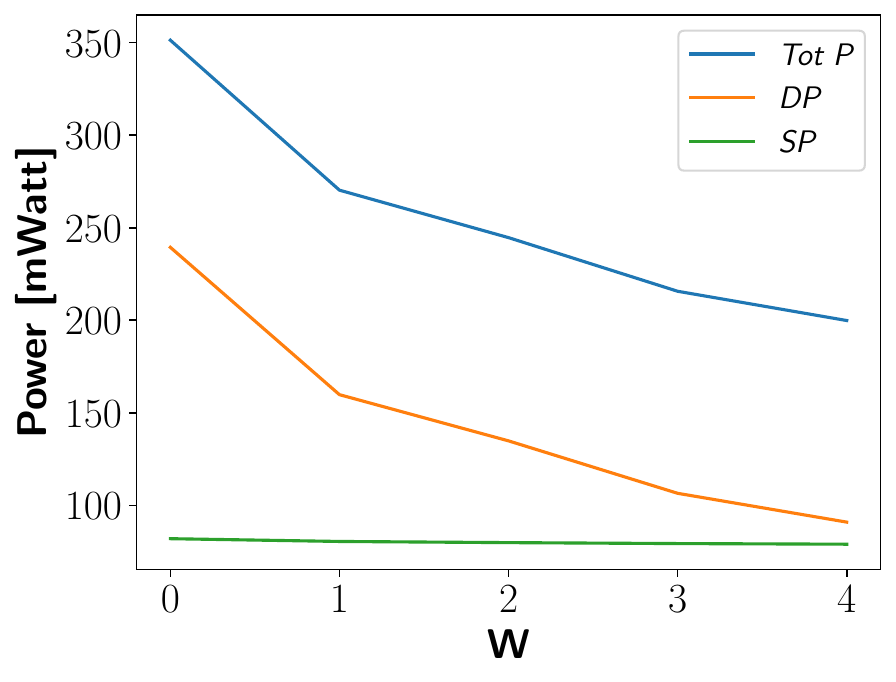}
	    \caption{Total (Tot P), static (SP) and dynamic power (DP) consumed as a function of the windowing order ($W$), considering five qubits and Q equal to four}
	    \label{fig:PowerVaryingW}
	\end{subfigure}
	\begin{subfigure}[t]{\columnwidth}
	    \centering
	    \includegraphics[width=0.9\textwidth]{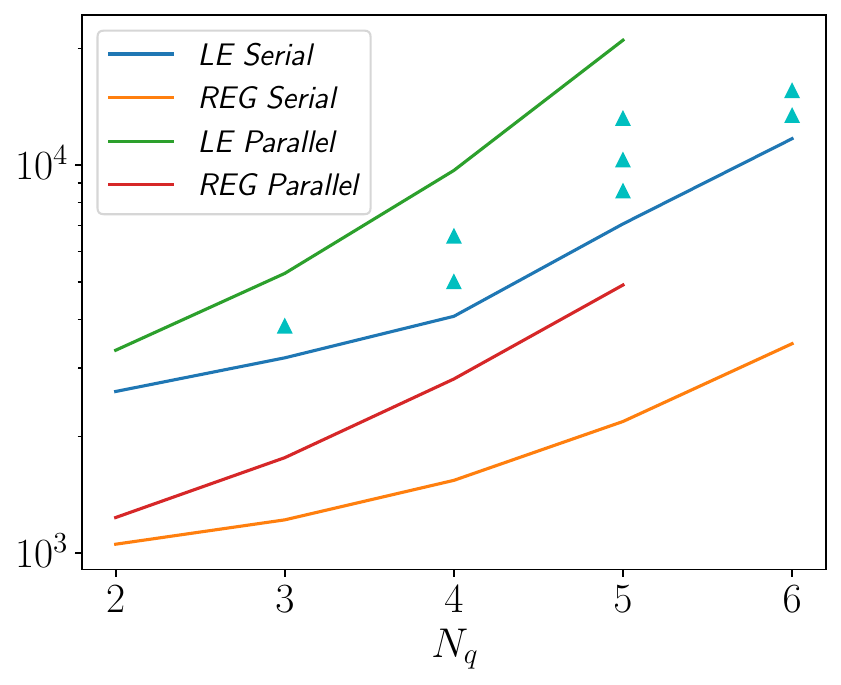}
	    \caption{Number of logic elements (LE) and registers (REG) required as a function of the number of qubits ($N_q$) in full serial and full parallel AEQUAM architecture, considering Q equal to four (cyano triangles represent the windowed intermediate versions)  }
	    \label{fig:AreaVaryingN}
	\end{subfigure}
        \quad
 	\begin{subfigure}[t]{\columnwidth}
	    \centering
\includegraphics[width=0.95\textwidth]{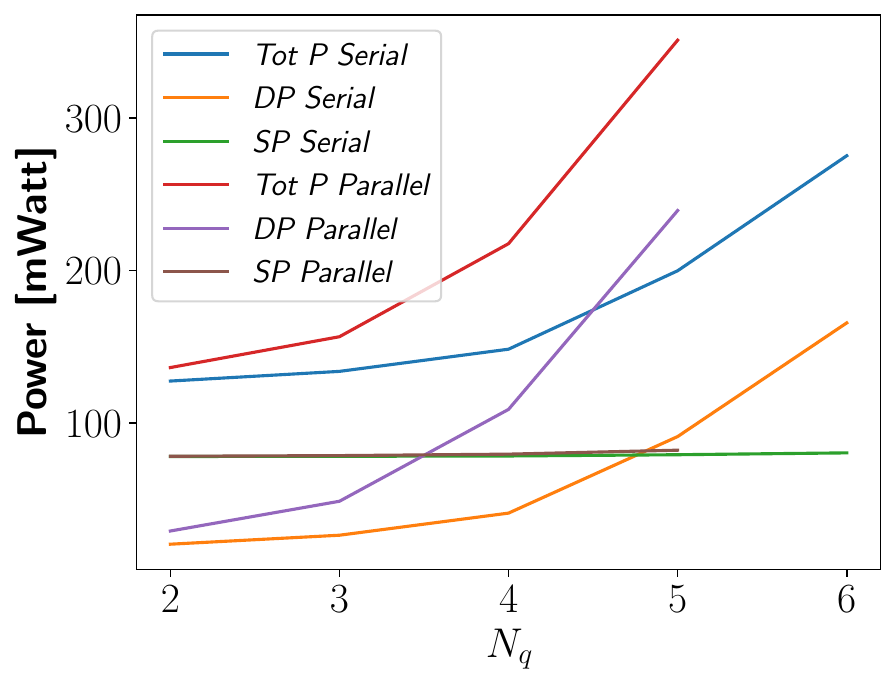}
	    \caption{Total (Tot P), static (SP) and dynamic power (DP) consumed as a function of the number of qubits ($N_q$) in full serial and full parallel AEQUAM architecture, considering Q equal to four (cyano triangles represent the windowed intermediate versions) }
	    \label{fig:PowerVaryingN}
	\end{subfigure}
	\caption{ Synthesis results obtained. }
	\label{fig:SynthesisResults}
\end{figure*}

\begin{table*}[htbp]
\centering
\caption{FPGA Resource Utilization for Different Configurations}
\label{tab:FPGAresources}
\begin{tabular}{cccccccc}
\toprule
\textbf{N} & \textbf{W} & \textbf{Q} & \textbf{LUTs} & \textbf{Registers} & \textbf{Total Power (mW)} & \textbf{Dynamic Power (mW)} & \textbf{Static Power (mW)} \\
\midrule
2 & 0 & 2 & 2489 & 752 & 130.41 & 23.15 & 77.89 \\
2 & 0 & 3 & 2743 & 914 & 133.42 & 26.14 & 77.91 \\
2 & 0 & 4 & 3335 & 1236 & 136.27 & 28.89 & 77.99 \\
2 & 0 & 5 & 4413 & 1878 & 139.75 & 32.31 & 78.10 \\
2 & 0 & 6 & 6645 & 3160 & 150.87 & 43.13 & 78.36 \\
2 & 1 & 2 & 2709 & 854 & 130.67 & 23.48 & 77.89 \\
2 & 1 & 3 & 3006 & 1041 & 133.89 & 26.48 & 77.91 \\
2 & 1 & 4 & 3628 & 1397 & 136.90 & 29.21 & 78.00 \\
2 & 1 & 5 & 4774 & 2161 & 140.46 & 32.63 & 78.10 \\
2 & 1 & 6 & 7238 & 3696 & 151.71 & 43.34 & 78.37 \\
\bottomrule
\end{tabular}
\end{table*}

\subsubsection{Discussion}
Figure \ref{fig:SynthesisResults} and Table \ref{tab:FPGAresources} present the results obtained by synthesizing the AEQUAM architecture on an \textbf{Intel Cyclone 10LP 10CL025YE144C8G} FPGA, which has a total Logic Element (LE) capacity of \textbf{24624}. The synthesis was performed by varying the architecture's degrees of freedom—specifically, the number of qubits ($N_q$), the windowing order ($W$), and the number of sine and cosine values that can be stored ($2^Q$).\\
In particular, Figures \ref{fig:AreaVaryingW} and \ref{fig:PowerVaryingW} display how area utilization and power consumption vary with $Q$, with $W$ set to zero and $N_q$ set to five. As anticipated, the value of $Q$ primarily affects the number of required registers, which shows an exponential increase as $Q$ increases.\\
On the other hand,  Figures \ref{fig:AreaVaryingW} and \ref{fig:PowerVaryingW} show the impact of varying the windowing order $W$ on area utilization and power consumption, with $Q$ fixed at four (i.e., sixteen sine and cosine values) and $N_q$ fixed at five. In this scenario, both area and power consumption decrease exponentially as $W$ increases. This behavior aligns with expectations, as the number of required datapaths in the windowed architecture is given by $\frac{2^{N_q-1}}{2^W}$.\\
Finally, Figures \ref{fig:AreaVaryingW} and \ref{fig:PowerVaryingW} depict the variation in area utilization and power consumption with respect to the number of qubits. This analysis is conducted for full serial, fully parallel, and windowed architectures (cyano triangles), with $Q$ set to four (corresponding to sixteen sine and cosine values). As expected, both area and power consumption increase \textbf{exponentially with the number of qubits}, as the number of probability amplitudes to handle grows as $2^{N_q}$. However, the significant area and power savings achieved through the windowing mechanism are evident.
\begin{table*}
\caption{Comparison between AEQUAM synthesis results and the current literature. }
\begin{center}
\begin{tabular}{?c?c?c?c?c?c?c?c?c?}
\noalign{\hrule height 1.5pt}
 \multirow{2}{*}{\textbf{Emulator}} & \multicolumn{2}{c?}{\textbf{AEQUAM}} & \multicolumn{4}{c?}{\textbf{Existing Emulators}} \\
\cline{2-7}
 & \textbf{Serial} & \textbf{Parallel} &\cite{pilch2019fpga} & \cite{mahmud2018scalable} & \cite{mahmud2020efficient}&\cite{Reis} \\
\noalign{\hrule height 1.5pt}
$N_{\textrm{qubit}}$ & 6 & 5 & 2 & 4 & 32   &  9 \\
\hline
\textbf{Devices} & Intel Cyclone 10LP  & Intel Cyclone 10LP   & Intel Cyclone V & Intel Arria 10 & Intel Arria 10 & Intel Stratix \\
\hline
\textbf{Logic Utilization} & 11702 LE  & 20993 LE & 8000 ALMs & 374021 ALMs & 56219 ALMs & 4019 LC \\
\hline
\textbf{Precision} &  20-bit fixed & 20-bit fixed & 10-bit fixed  & 32-bit floating & 64-bit floating & 18-bit fixed \\
\hline
\textbf{Clock Frequency} & \SI{80}{\mega \hertz} & \SI{109}{\mega \hertz}  & - & \SI{233}{\mega \hertz}& \SI{233}{\mega \hertz} & - \\
\noalign{\hrule height 1.5pt}
\end{tabular}
\end{center}
\label{tab:TableComparisons}
\end{table*}
\begin{figure}
	\centering	\includegraphics[width=0.5\textwidth]{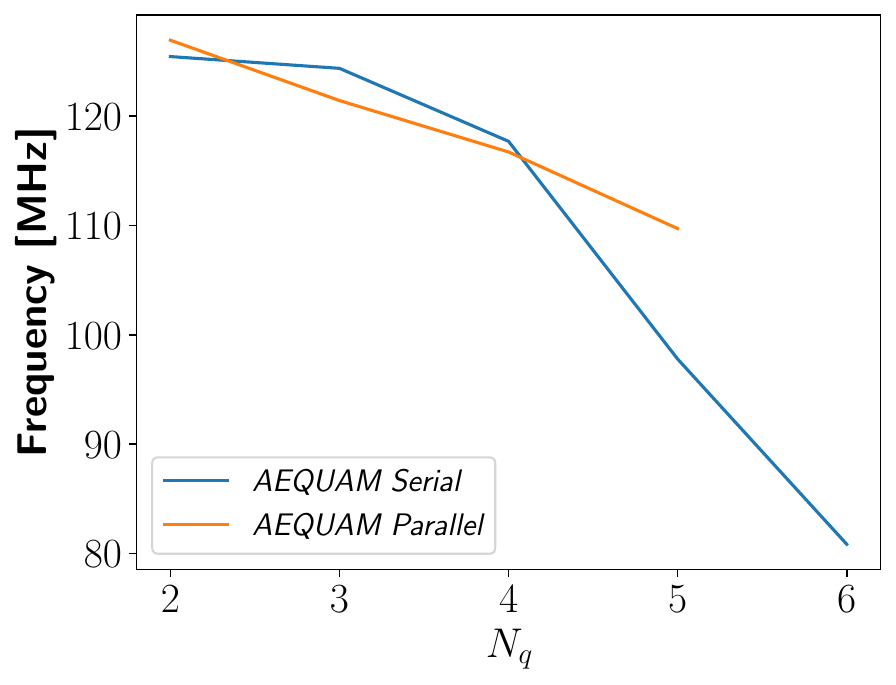}
\caption{ Maximum operating frequency as a function of the number of qubits ($N_q$) in full serial and full parallel AEQUAM architecture, considering Q equal to four .}
\label{fig:Frequency}
\end{figure}\begin{figure}
	\centering	\includegraphics[width=0.5\textwidth]{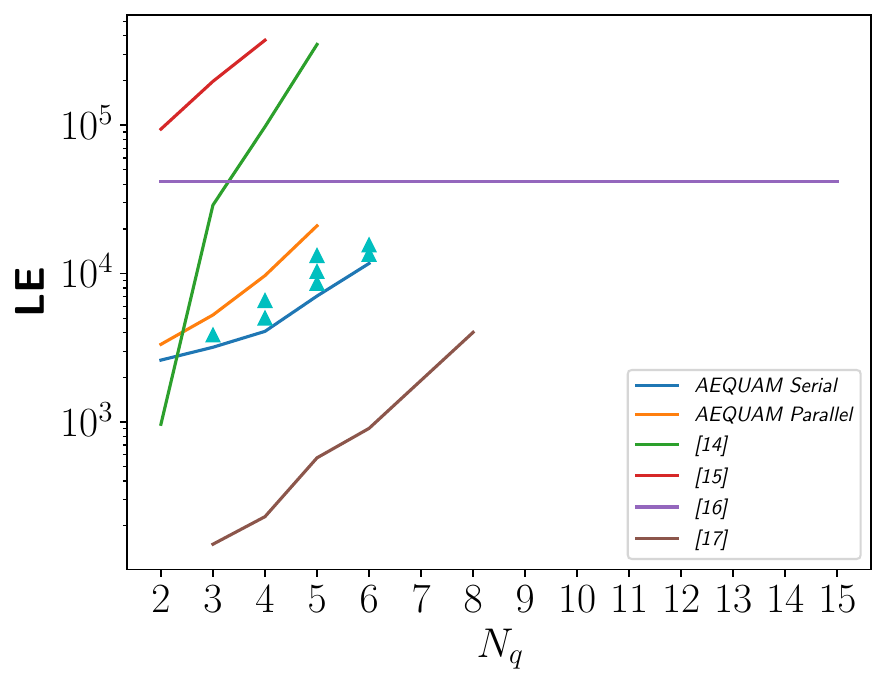}
\caption{ Number of logic elements (LE) required as a function of the number of qubits ($N_q$) in AEQUAM architecture (serial, parallel and windowed versions) and the state-of-the-art emulators. }
\label{fig:Comparison}
\end{figure}\\
Figure \ref{fig:Frequency} illustrates the maximum operating frequency achieved with both serial and parallel AEQUAM architectures as the number of qubits varies. These frequency variations are primarily influenced by the size of the selection and reordering units, whose depth increases nearly linearly with the number of qubits involved. This increase in depth leads to a corresponding linear decrease in the maximum operating frequency. It is important to note that the current architecture is not pipelined; the future addition of pipeline stages is expected to positively impact the operating frequency.
\\
The obtained results are compared with the state-of-the-art emulators in Figure \ref{fig:Comparison} and Table \ref{tab:TableComparisons}. Figure \ref{fig:Comparison} shows that the architecture proposed by \cite{Reis} is the only one with better scalability than ours. However, it is essential to note that the architecture in \cite{Reis} \textbf{ does not support rotational gates}, which are essential for executing many quantum circuits. Additionally, \cite{mahmud2020efficient} shows a flat scalability curve because, in this architecture, the probability amplitudes are stored in external memory, and the computation is not parallelized. Moreover, this architecture is specialized for a single quantum circuit, i.e. quantum fast Fourier transform. Both the table and the plot demonstrate that \textbf{the proposed architecture is competitive} with the current state of the art, offering good scalability without the use of external memories and supporting both Clifford+T and rotational gates, thereby enabling the execution of any generic quantum circuit.

\section{Discussion}\label{sec:discussion}
Simulation on classical hardware, particularly when performed with hardware-aware tools such as AEQUAM, remains the most accessible and scalable method for quantum algorithm prototyping. Indeed, it allows developers to explore algorithm structure, resource requirements, and hardware-software integration challenges without access to expensive or noisy quantum hardware.\\
The results obtained from both software and hardware implementations prove the AEQUAM’s effectiveness in reducing execution time while maintaining a high level of accuracy.\\
The floating-point software model provides a highly accurate reference but is computationally expensive, requiring exponential time and memory. AEQUAM’s fixed-point FPGA implementation introduces negligible approximation errors but enables significantly reduced hardware resource usage.\\
Unlike previous FPGA-based quantum emulators that are based on direct matrix-vector multiplication, AEQUAM exploits a sparsity-aware architecture based on a butterfly-like selection mechanism. This approach enables more efficient use of logic elements (LEs), allowing AEQUAM to scale better than existing FPGA-based implementations. It is possible to notice that while architectures such as \cite{pilch2019fpga} achieve only two-qubit emulation on a Cyclone V FPGA, AEQUAM demonstrates six-qubit emulation on a smaller Intel Cyclone 10LP FPGA, showcasing better area efficiency.\\
The experiments confirm that 20-bit fixed-point representation with nearest rounding achieves an optimal balance between accuracy and FPGA resource efficiency. While floating-point implementations offer marginally higher fidelity, they consume substantially more area and introduce unnecessary complexity. Moreover, as shown in Figure \ref{fig:ResultsPlot}, AEQUAM achieves similar fidelity to floating-point software models while dramatically reducing execution time.\\
These results highlight AEQUAM’s potential for real-time quantum circuit validation and scalability for larger quantum simulations.

\section{Conclusions}\label{sec:conclusions}
This article has introduced the \textbf{AEQUAM} toolchain, designed to accelerate and simplify the validation of quantum algorithms. By implementing a hardware emulator with configurable precision and parallelization, AEQUAM provides a scalable alternative to traditional software-based simulators.\\
Key conclusions of this work include:
\begin{itemize}
    \item Significant reduction in execution time compared to software emulators, achieved through parallelized gate execution and sparsity-aware computation.
    \item Competitive scalability, demonstrating six-qubit emulation on an Intel Cyclone 10LP FPGA, outperforming prior FPGA-based solutions in terms of resource efficiency.
    \item Configurable numerical precision, allowing users to balance accuracy and hardware complexity for different quantum circuits.
\end{itemize}
Although the current version of AEQUAM has demonstrated the potential of this approach with promising results, there is considerable room for further enhancement to address its limitations and expand its applicability. Future work could focus on developing efficient hardware modules for performing trigonometric operations directly on board, eliminating the need for precomputed values and reducing memory overhead. Additionally, reorganizing the datapath to implement a pipelined architecture would allow better exploitation of the parallelizability of probability amplitude pairs without requiring unit replication. This improvement could significantly mitigate the delay in sequential or partially sequential emulators.\\
Extending compiler support to include the openQASM 3.0 specification would enable AEQUAM to handle more advanced quantum circuits and dynamic constructs, such as loops and conditionals, thereby improving its compatibility with state-of-the-art quantum frameworks. Furthermore, enhancing memory management strategies—such as optimizing data allocation and retrieval processes—would improve the architecture’s scalability, enabling it to emulate circuits with a larger number of qubits and gates on resource-constrained FPGA platforms. Exploring alternative number representations, such as custom floating-point formats that maintain high accuracy while using fewer bits, could also contribute to this goal.       \\
Finally, including support for real-time measurements could further broaden AEQUAM’s utility, making it a versatile tool for both research and practical applications in quantum computing.\\
In conclusion, the AEQUAM toolchain marks a significant step forward for efficient quantum algorithm validation, offering a practical and scalable solution. As quantum computing evolves, AEQUAM stays ahead, continually adapting to new and complex quantum circuits, empowering researchers and developers to push the boundaries of innovation.

\bibliographystyle{ieeetr}
\bibliography{sn-bibliography}

\begin{thebibliography}{10}

\bibitem{comboptsurvey}
E.~Zahedinejad and A.~Zaribafiyan, ``Combinatorial optimization on gate model quantum computers: A survey,'' 2017.
\newblock \url{https://doi.org/10.48550/arXiv.1708.05294}.

\bibitem{comboptimizationquantumvariational}
C.~Grange, M.~Poss, and E.~Bourreau, ``An introduction to variational quantum algorithms on gate-based quantum computing for combinatorial optimization problems,'' 2022.
\newblock \url{https://doi.org/10.48550/arxiv.2212.11734}.

\bibitem{GroverPatternPerformance}
Y.~Liu and G.~J. Koehler, ``Using modifications to grover’s search algorithm for quantum global optimization,'' {\em European Journal of Operational Research}, vol.~207, no.~2, pp.~620--632, 2010.
\newblock \url{https://doi.org/10.1016/j.ejor.2010.05.039}.

\bibitem{gaspapervlsi}
L.~Giuffrida, D.~Volpe, G.~A. Cirillo, M.~Zamboni, and G.~Turvani, ``Engineering grover adaptive search: Exploring the degrees of freedom for efficient qubo solving,'' {\em IEEE Journal on Emerging and Selected Topics in Circuits and Systems}, vol.~12, no.~3, pp.~614--623, 2022.
\newblock \url{https://doi.org/10.1109/JETCAS.2022.3202566}.

\bibitem{biamonte2017quantum}
J.~Biamonte, P.~Wittek, N.~Pancotti, P.~Rebentrost, N.~Wiebe, and S.~Lloyd, ``Quantum machine learning,'' {\em Nature}, vol.~549, no.~7671, pp.~195--202, 2017.
\newblock \url{https://doi.org/10.1038/nature23474}.

\bibitem{qmlsurvey}
N.~Mishra, M.~Kapil, H.~Rakesh, A.~Anand, N.~Mishra, {\em et~al.}, ``Quantum machine learning: A review and current status,'' in {\em Data Management, Analytics and Innovation}, (Singapore), pp.~101--145, Springer Singapore, 2021.
\newblock \url{https://doi.org/10.1007/978-981-15-5619-7_8}.

\bibitem{quantumchemistry}
Y.~Cao, J.~Romero, J.~P. Olson, M.~Degroote, P.~D. Johnson, M.~Kieferová, I.~D. Kivlichan, T.~Menke, B.~Peropadre, N.~P.~D. Sawaya, S.~Sim, L.~Veis, and A.~Aspuru-Guzik, ``Quantum chemistry in the age of quantum computing,'' Aug 2019.
\newblock \url{https://doi.org/10.1021/acs.chemrev.8b00803}.

\bibitem{pirnay2022super}
N.~Pirnay, V.~Ulitzsch, F.~Wilde, J.~Eisert, and J.-P. Seifert, ``A super-polynomial quantum advantage for combinatorial optimization problems,'' {\em arXiv preprint arXiv:2212.08678}, 2022.
\newblock \url{ https://doi.org/10.48550/arXiv.2212.08678}.

\bibitem{riste2017demonstration}
D.~Rist{\`e}, M.~P. Da~Silva, C.~A. Ryan, A.~W. Cross, A.~D. C{\'o}rcoles, J.~A. Smolin, J.~M. Gambetta, J.~M. Chow, and B.~R. Johnson, ``Demonstration of quantum advantage in machine learning,'' {\em npj Quantum Information}, vol.~3, no.~1, p.~16, 2017.
\newblock \url{https://doi.org/10.1038/s41534-017-0017-3}.

\bibitem{yuan2020quantum}
X.~Yuan, ``A quantum-computing advantage for chemistry,'' {\em Science}, vol.~369, no.~6507, pp.~1054--1055, 2020.
\newblock \url{https://doi.org/10.1126/science.abd3880}.

\bibitem{surveyEmulation}
H.~Li and Y.~Pang, ``Fpga-accelerated quantum computing emulation and quantum key distillation,'' {\em IEEE Micro}, vol.~41, no.~4, pp.~49--57, 2021.

\bibitem{memBottleneck}
X.-C. Wu, S.~Di, E.~M. Dasgupta, F.~Cappello, H.~Finkel, Y.~Alexeev, and F.~T. Chong, ``Full-state quantum circuit simulation by using data compression,'' in {\em Proceedings of the International Conference for High Performance Computing, Networking, Storage and Analysis}, p.~1–24, ACM, Nov. 2019.
\newblock \url{https://doi.org/10.1145/3295500.3356155}.

\bibitem{openqasm2}
A.~W. Cross, L.~S. Bishop, J.~A. Smolin, and J.~M. Gambetta, ``Open quantum assembly language,'' 2017.
\newblock \url{https://doi.org/10.48550/ARXIV.1707.03429}.

\bibitem{nielsen_quantum_2010}
M.~A. Nielsen and I.~L. Chuang, {\em Quantum computation and quantum information}.
\newblock Cambridge ; New York: Cambridge University Press, 10th anniversary ed~ed., 2010.

\bibitem{pilch2019fpga}
J.~Pilch and J.~D{\l}ugopolski, ``An fpga-based real quantum computer emulator,'' {\em Journal of Computational Electronics}, vol.~18, pp.~329--342, 2019.
\newblock \url{https://doi.org/10.1007/s10825-018-1287-5}.

\bibitem{mahmud2018scalable}
N.~Mahmud and E.~El-Araby, ``A scalable high-precision and high-throughput architecture for emulation of quantum algorithms,'' in {\em 2018 31st IEEE International System-on-Chip Conference (SOCC)}, pp.~206--212, IEEE, 2018.
\newblock \url{https://doi.org/10.1109/SOCC.2018.8618545}.

\bibitem{mahmud2020efficient}
N.~Mahmud, B.~Haase-Divine, A.~Kuhnke, A.~Rai, A.~MacGillivray, and E.~El-Araby, ``Efficient computation techniques and hardware architectures for unitary transformations in support of quantum algorithm emulation,'' {\em Journal of Signal Processing Systems}, vol.~92, pp.~1017--1037, 2020.
\newblock \url{https://doi.org/10.1007/s11265-020-01569-4}.

\bibitem{Reis}
C.~Concei{\c{c}}{\~a}o and R.~Reis, ``Efficient emulation of quantum circuits on classical hardware,'' in {\em 2015 IEEE 6th Latin American Symposium on Circuits \& Systems (LASCAS)}, pp.~1--4, IEEE, 2015.
\newblock \url{https://doi.org/10.1109/LASCAS.2015.7250404}.

\bibitem{negovetic2002evolving}
G.~Negovetic, M.~Perkowski, M.~Lukac, and A.~Buller, ``Evolving quantum circuits and an fpga-based quantum computing emulator,'' 2002.
\newblock \url{https://pdxscholar.library.pdx.edu/ece_fac/191/}.

\bibitem{nussbaumer1982fast}
H.~J. Nussbaumer and H.~J. Nussbaumer, {\em The fast Fourier transform}.
\newblock Springer, 1982.
\newblock \url{https://doi.org/10.1007/978-3-642-81897-4_4}.

\bibitem{OPENQASM}
A.~W. Cross, L.~S. Bishop, J.~A. Smolin, and J.~M. Gambetta, ``Open quantum assembly language,'' 2017.
\newblock \url{https://doi.org/10.48550/ARXIV.1707.03429}.

\bibitem{BAKIRI2018135}
M.~Bakiri, C.~Guyeux, J.-F. Couchot, and A.~K. Oudjida, ``Survey on hardware implementation of random number generators on fpga: Theory and experimental analyses,'' {\em Computer Science Review}, vol.~27, pp.~135--153, 2018.
\newblock \url{https://doi.org/10.1016/j.cosrev.2018.01.002}.

\bibitem{koren2018computer}
I.~Koren, {\em Computer arithmetic algorithms}.
\newblock CRC Press, 2018.

\bibitem{virtlab}
M.~Ruo~Roch and M.~Martina, ``Virtlab: A low-cost platform for electronics lab experiments,'' {\em Sensors}, vol.~22, no.~13, 2022.
\newblock \url{https://doi.org/10.3390/s22134840}.

\bibitem{quetschlich2023mqt}
N.~Quetschlich, L.~Burgholzer, and R.~Wille, ``Mqt bench: Benchmarking software and design automation tools for quantum computing,'' {\em Quantum}, vol.~7, p.~1062, 2023.
\newblock \url{ https://doi.org/10.22331/q-2023-07-20-1062}.

\bibitem{processorservervlsi}
``Intel {X}eon {G}old 6134 processor - product specification.''
\newblock [Online] \url{https://ark.intel.com/content/www/us/en/ark/products/120493/intel-xeon-gold-6134-processor-24-75m-cache-3-20-ghz.html}, accessed 25-October-2021.

\bibitem{noauthor_qiskitquantum_infohellinger_fidelity_nodate}
``qiskit.quantum\_info.hellinger\_fidelity — {Qiskit} 0.36.2 documentation.''
\newblock \url{https://tinyurl.com/mr27xdmc}.

\bibitem{kullback_information_1951}
S.~Kullback and R.~A. Leibler, ``On {Information} and {Sufficiency},'' {\em The Annals of Mathematical Statistics}, vol.~22, pp.~79--86, Mar. 1951.
\newblock \url{https://doi.org/10.1214/aoms/1177729694}.

\end{thebibliography}


\EOD
\begin{IEEEbiography}[{\includegraphics[width=1in,height=1.30in,clip,keepaspectratio]{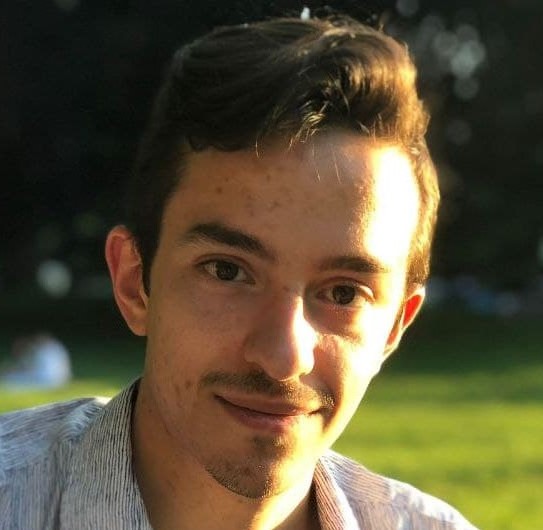}}]{Lorenzo Lagostina} received the electronics and communications engineering from the Politecnico di Torino, Torino, Italy, in 2022. He is currently a Ph.D. student in the Department of Electronics and Telecommunications of the Politecnico di Torino. His research interests include Electronic Design Automation (EDA), and Embedded Machine Learning. In 2022 he received the second prize in the IEEE R8 Student Paper Contest at the MELECON22 conference.
\end{IEEEbiography}

\begin{IEEEbiography}[{\includegraphics[width=1in,height=1.30in,clip,keepaspectratio]{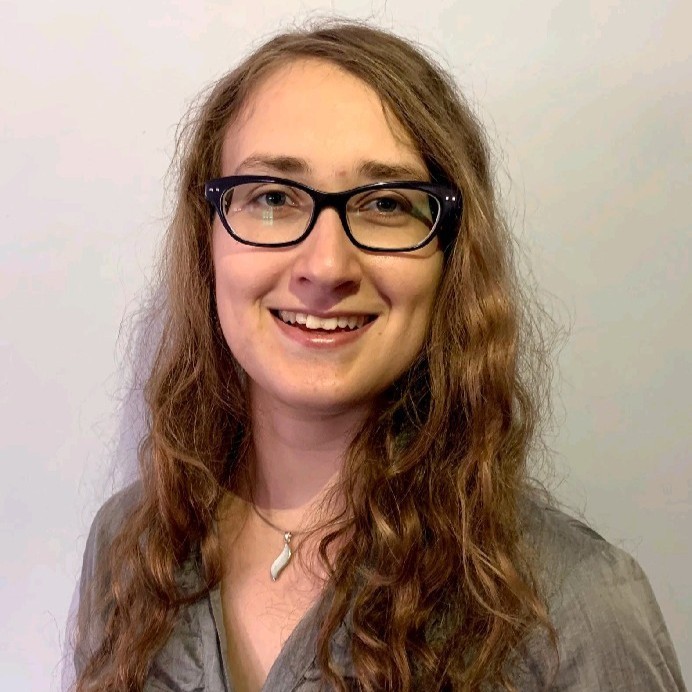}}]{Deborah Volpe} (Graduate Student Member, IEEE) received the B.Sc. and M.Sc. degrees in Electronic Engineering – in 2019 and 2021, respectively – from Politecnico di Torino, where she is now pursuing the Ph.D. degree in Electrical, Electronics and Communications Engineering. Her research interests mainly focus on the emulation of quantum computers on classical hardware (FPGA, CPU and GPU) and quantum-compliant approaches for solving QUBO problems.
\end{IEEEbiography}

\begin{IEEEbiography}[{\includegraphics[width=1in,height=1.3in,clip,keepaspectratio]{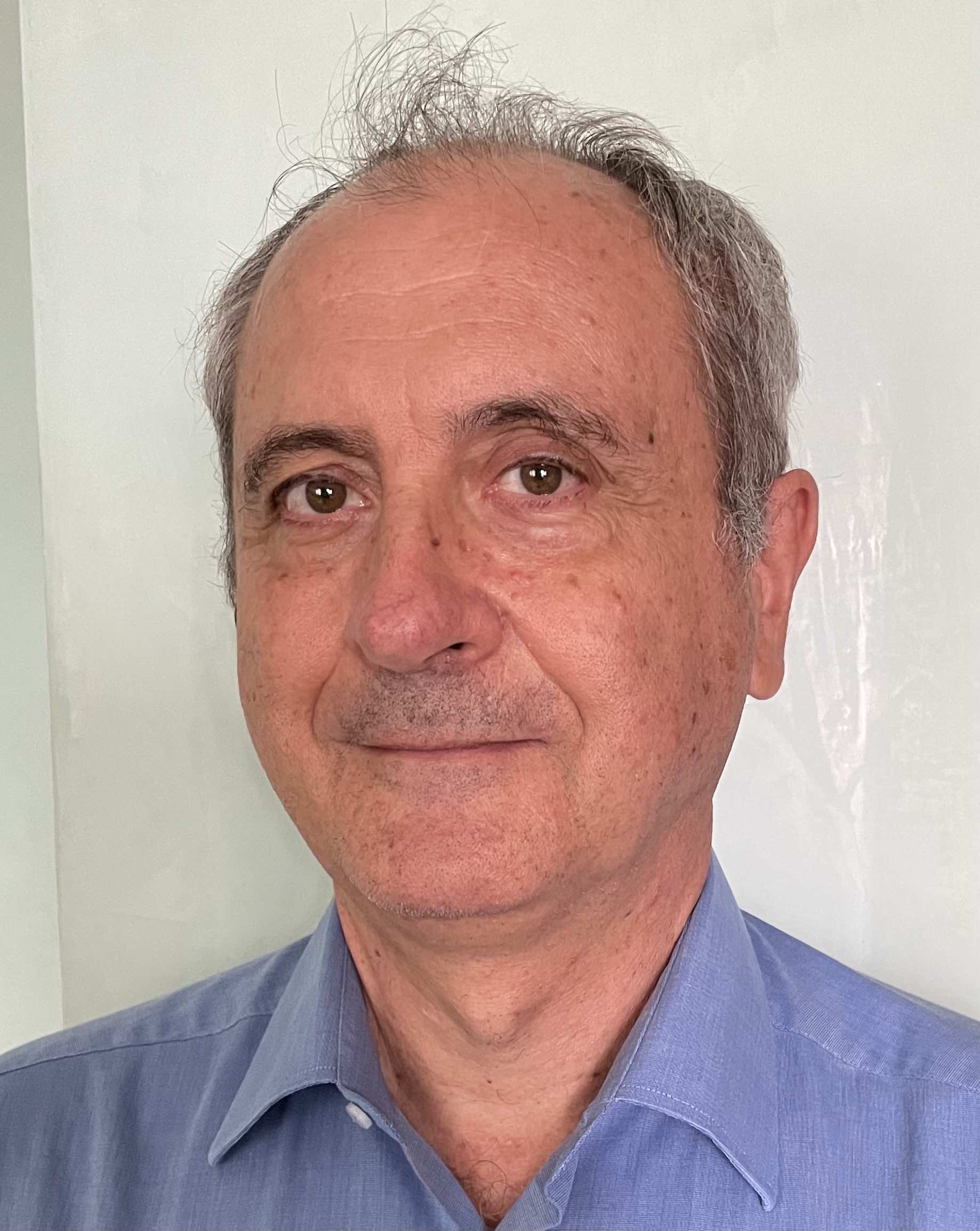}}]{Maurizio Zamboni}received the Degree in   Electronics Engineering  in 
1983 and the Ph. D. degree in  1988 at the Politecnico di Torino. He joined the Electronics Department, Politecnico di Torino, in 1983, became Researcher in 1989, Associate Professor in
1992 and  Full Professor of Electronics in 2005.
His research activity started with the study of multiprocessor architectures, then he worked with digital IC design concentrating both on architectural aspects and circuits optimisation.
In these years he got an expertise in the design of special ICs for Artificial Intelligence, Vision and Telecommunication.
His main interests include now low-power circuits and innovative technologies  beyond the CMOS world such as Nano Magnetic Logic and NanoArrays.
 From 2018 he started moving to Quantum Computing issues, contributing to the creation, at Politecnico of Torino,  of a research group very active in many fields of the QC world, from technologies to emulators/simulators up to AI and ML applications. Many activities have been carried on from the device modeling for the promising technologies, up to the development of basic cells models for Quantum Gate Arrays and the study of applications of QC in the world of Communications, Security and AI.
He is co-author of more than 200 scientific papers (three invited papers) and three books and holds two patents.
\end{IEEEbiography}

\begin{IEEEbiography}[{\includegraphics[width=1in,height=1.3in,clip,keepaspectratio]{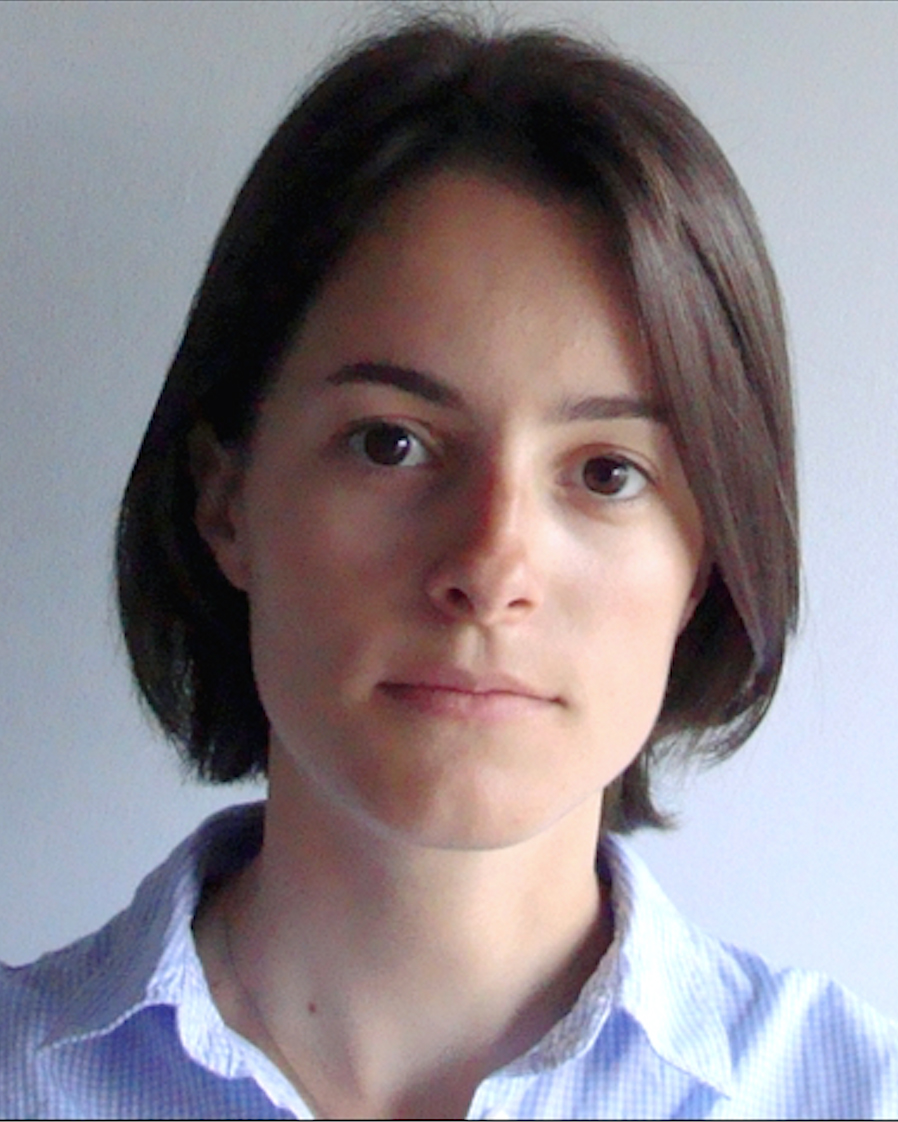}}]{Giovanna Turvani}received the M.Sc. degree with honours (Magna Cum Laude) in Electronic Engineering in 2012 and the Ph.D. degree from the Politecnico di Torino. She was Postdoctoral Research Associate at the Technical University of Munich in 2016. She is currently Assistant Professor at Politecnico di Torino. Her interests include CAD Tools development for non-CMOS nanocomputing, architectural design for field-coupled nanocomputing and high-level device modelling for Quantum Computing and hardware systems for microwave imaging-based techniques for biomedical applications and for food quality monitoring. Other expertise includes also the design of IoT low-power systems based on long-range protocols (LoRa).
\end{IEEEbiography}

\end{document}